\documentclass[aps,preprint,nofootinbib,preprintnumbers]{revtex4}

\usepackage{graphicx,amsmath,amssymb,bm,subfigure,comment}

\renewcommand\d{\partial}
\newcommand\<{\langle}
\renewcommand\>{\rangle}
\newcommand\+{\dagger}
\newcommand\p{{\mathbf{p}}}
\newcommand\q{{\mathbf{q}}}
\newcommand\x{{\mathbf{x}}}
\newcommand\y{{\mathbf{y}}}
\newcommand\Tr{{\mathop{\mathrm{Tr}}}}

\newcommand\beq{\begin{eqnarray}}
\newcommand\eeq{\end{eqnarray}} 
\newcommand\eqn[1]{\label{eq:#1}} 
\newcommand\eq[1]{eq.~(\ref{eq:#1})} 
\newcommand\Ro{{R_0}}
\newcommand\Rp{{R_+}}
\newcommand\Rm{{R_-}}
\newcommand\Rpm{{R_\pm}}

\begin{document}

\preprint{INT-PUB-09-020}
\title{Conformality Lost}
\author{David~B.~Kaplan}
\affiliation{Institute for Nuclear Theory, University of Washington,
Seattle, Washington 98195-1550, USA}
\author{Jong-Wan Lee}
\affiliation{Institute for Nuclear Theory, University of Washington,
Seattle, Washington 98195-1550, USA}
\author{Dam T.~Son}
\affiliation{Institute for Nuclear Theory, University of Washington,
Seattle, Washington 98195-1550, USA}
\author{Mikhail A.~Stephanov}
\affiliation{Department of Physics, University of Illinois, Chicago, 
Illinois 60607-7059, USA}
\date{May 2009}

\begin{abstract}

We consider zero-temperature transitions from conformal to
non-conformal phases in quantum theories.  We argue that there are
three generic mechanisms for the loss of conformality in any number of
dimensions: (i) fixed point goes to zero coupling, (ii) fixed point
runs off to infinite coupling, or (iii) an IR  fixed point
annihilates with a UV fixed point and they both disappear into
the complex plane.  We give both relativistic and non-relativistic
examples of the last case in various dimensions and show that the
critical behavior of the mass gap behaves similarly to the correlation
length in the finite temperature Berezinskii-Kosterlitz-Thouless (BKT)
phase transition in two dimensions, $\xi \sim {\rm exp}(c/
|T-T_c|^{1/2})$. We speculate that the chiral phase transition in QCD
at large number of fermion flavors belongs to this universality class,
and attempt to identify the UV fixed point that annihilates with the
Banks-Zaks fixed point at the lower end of the conformal window.

\end{abstract}
%\keywords{Berezinskii-Kosterlitz-Thouless phase transition, 
%Breitenlohner-Freedman bound, Efimov effect, Miransky scaling}
\maketitle

\section{Introduction}

The renormalization group (RG) underlies our understanding of
second-order phase transitions, with critical points being identified
with fixed points of the appropriate RG equation~\cite{Wilson:1973jj}.
%In many cases critical points for a classical theory in $d$ spatial
%dimensions at finite temperature are governed by the fixed point in a
%zero temperature quantum field theory in $d+1$ spacetime dimensions.
Near the phase transition the characteristic energy or momentum scale
$m$ (the inverse correlation length) goes to zero as $m \sim
|\alpha-\alpha_*|^\nu$, where $\alpha$ is a parameter that can vary
continuously, and $\alpha=\alpha_*$ is the location of the critical
point.

In this paper, we argue that there is wide class of phase transitions
in which the correlation length behaves very differently, vanishing
exponentially on one side of the phase transition, while being
strictly zero on the other side
\beq
\eqn{BKT}
  m 
  \sim
  \Lambda_\text{UV}\, \theta(\alpha_*-\alpha)\,
\exp \left(-\displaystyle{\frac c{\sqrt{\alpha_*-\alpha}}}\right) ,
\qquad c>0\ .
\eeq
This peculiar behavior --- where all derivatives of the correlation
length with respect to $\alpha$ vanish at the critical point --- has
been observed before in the Berezinskii-Kosterlitz-Thouless (BKT)
phase transition in two dimensions~\cite{Kosterlitz}; therefore we
will refer to \eq{BKT} as ``BKT scaling.''  The BKT transition is a
classical phase transition in two dimensions that can be described in
terms of vortex condensation.  It arises due to the competition
between the entropy of a single vortex and the binding energy of a
pair of vortices, both which scale as $\log R$, $R$ being the size of
the system.  While this transition is peculiar to two dimensions, we
will show that the mechanism underlying BKT scaling from an RG point
of view is far more general, and is one of three generic behaviors 
that can occur when a system in any dimension makes a transition from a
conformal to a non-conformal phase.  In particular, as we will show, 
it follows when an
IR fixed point of the system merges with a UV fixed
point.  In this language it is easy to see why BKT scaling can be
found in a wide variety of systems.

\begin{figure}[ht]
\centering
\subfigure[]{
\includegraphics[width=0.40\textwidth]{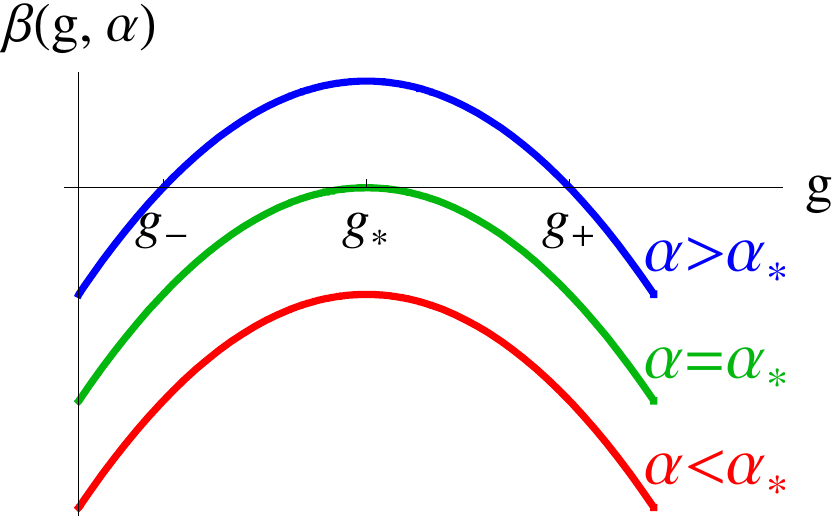}
\label{fig:rg}
}
\ \ \ \ \ \ \ 
\subfigure[]{
\includegraphics[width=0.40\textwidth]{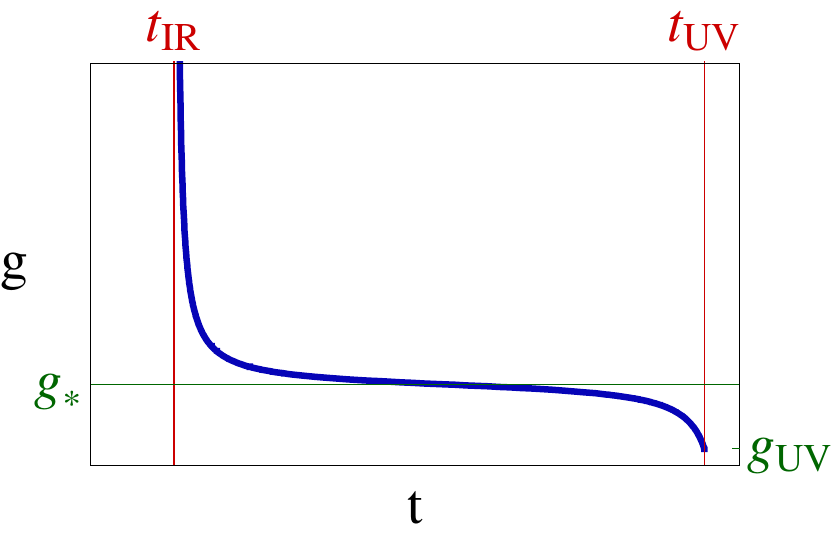}
\label{fig:gflow}
}
\caption{(a) A toy $\beta$-function. For $\alpha>\alpha_*$ there are fixed
points at $g_\pm$ which are UV- and IR-stable respectively; these
fixed points merge at $g_*$ for $\alpha=\alpha_*$, and disappear for
$\alpha<\alpha_*$; (b) The RG flow of the coupling $g$ as a function
of $t=\ln\mu$ in the non-conformal phase, with
$(t_\text{UV}-t_\text{IR})\propto 1/\sqrt{\alpha_*-\alpha}$.}
\end{figure}

The basic mechanism can be illustrated with a simple model with a 
dimensionless coupling $g$ depending on an external
parameter $\alpha$, for which the $\beta$-function takes the 
form (Fig.~\ref{fig:rg})
\beq
\beta(g;\alpha) =\frac{\d g}{\d t} = (\alpha-\alpha_*) -  (g-g_*)^2\ ,
\eqn{toy}
\eeq
where $t=\ln\mu$, $\mu$ being the renormalization scale.  For
$(\alpha-\alpha_*)>0$, the fixed points for this system (zeros of
$\beta$) are given by
\beq
g_\pm = g_* \pm \sqrt{\alpha-\alpha_*}\ ,
\eqn{gtoy}
\eeq
where $g_-$, $g_+$ correspond to IR and UV fixed points
respectively, each describing a conformal phase of the 
theory\footnote{By IR and UV fixed points we mean zeros of the 
$\beta$-function which are attractive or repulsive in the IR respectively.}.  
As
$\alpha$ decreases, these two fixed points approach each other until
they merge at $g_\pm = g_*$ for $\alpha=\alpha_*$.  For $\alpha <
\alpha_*$ the solutions to $\beta=0$ are complex, and the theory no
longer has a conformal phase.

%%%%%%%%
%%%%%%%%%%%

To see that fixed point merger generically gives rise to BKT scaling,
consider the case where $\alpha$ is slightly below $\alpha_*$, and
that at a UV scale $\Lambda_\text{UV}$ the coupling takes an initial
value $g_\text{UV} < g_*$.  On scaling to the IR, the coupling then
flows to larger values, lingering near $g=g_*$ where the
$\beta$-function is small, and then blowing up quickly, defining an
intrinsic IR scale $\Lambda_\text{IR}$, which is insensitive to the
initial value $g_\text{UV}$.  This behavior is displayed in
Fig.~\ref{fig:gflow}.
The scale $\Lambda_\text{IR}$ will characterize the longest
correlation lengths in this theory, and can be computed by integrating
\eq{toy}:
\beq
\frac{\Lambda_\text{IR}}{  \Lambda_\text{UV}}
  =\exp\left[t_\text{IR}-t_\text{UV}\right]
  =\exp\left[\int_{g_\text{UV}}^{g_\text{IR}}\frac{dg}{\beta(g;\alpha)}\right]
  \simeq e^{-\pi/\sqrt{(\alpha_*-\alpha)}}\ ,
\eqn{toycorr}\eeq
where we have assumed  $\vert
g_\text{IR,UV}-g_*\vert \gg \vert\alpha-\alpha_*\vert$ .

How general is this mechanism of fixed point annihilation?  Suppose a
system has a nontrivial IR fixed point at $g=\bar g(\alpha)$ whose
location depends continuously on a parameter $\alpha$, and that at a
critical value $\alpha=\alpha_*$ there is a phase transition where
conformality is lost.  At this phase transition, the $\beta$-function
must somehow lose a zero.  This can come about in three ways:
\begin{enumerate}
\item[(A)] $\bar g$ can decrease until it merges with the trivial
fixed point at $g=0$, giving rise to a trivial, asymptotically unfree
theory;
\item[(B)] $\bar g$ can run off to infinite coupling and disappear;
%giving rise to a free magnetic phase;
\item[(C)] $\bar g$ can merge with a UV fixed point, as in our toy
model, giving rise to BKT scaling.
\end{enumerate}
Examples of scenarios (A) and (B) are afforded by supersymmetric QCD
(SQCD).  At large number of colors $N_c$, the parameter $x\equiv
N_f/N_c$ may be treated as continuous, where $N_f$ is the number of
quark flavors.  It has been shown by
Seiberg~\cite{Seiberg:1994pq,Intriligator:1995au} that SQCD is
conformal in the window $3/2 \le x\le 3$.  For $x$ just
below 3, the theory has a Banks-Zaks fixed point at weak coupling
\cite{Banks:1981nn}; approaching $x=3$ from below, this fixed
point merges with the trivial fixed point at $g=0$, and for $x>3$
the theory is in the asymptotically unfree ``free electric phase.''
This is an example of mechanism ``A'' above.  In contrast, at the
lower end of the conformal window at $x=3/2$, SQCD goes from a
strongly coupled conformal theory when $x\gtrsim 3/2$ to a ``free
magnetic phase'' when $x\lesssim 3/2$.  In the free magnetic
phase, the Coulomb force between charges takes the form $e^2 \ln
(\Lambda r)/r^2$ where $\Lambda$ is associated with the Landau pole of
the dual magnetic theory.  The log behavior of the coupling can be
explained by a $\beta$-function which is negative and approaches zero
as $\beta\sim -1/g$ for large $g$.  
%We take this to mean that
%conformality is lost via mechanism (B).
% and conversely that (B) implies a free magnetic phase.
Thus it appears that conformality in the electric description is lost via
mechanism (B).  [Yet, since in the dual magnetic theory conformality is lost
via mechanism (A), it would appear that scenarios (A) and (B) can describe 
the same physics in terms of different degrees of freedom.] 

In this paper we give several examples of theories which exhibit the
mechanism (C) of fixed point merger and BKT scaling.  Following our RG
analysis of the original BKT transition, we analyze the quantum
mechanical example of a $1/r^2$ potential in $d$ dimensions, which can
be solved nonperturbatively and which exhibits the phenomenon of fixed
point merger.  We show how this analysis has many parallels in the
AdS/CFT
correspondence~\cite{Maldacena:1997re,Gubser:1998bc,Witten:1998qj},
and that loss of conformality via fixed point merger is analogous 
(if not holographically dual) to
the instability of AdS space at the Breitenlohner-Freedman (BF) bound
\cite{Breitenlohner:1982jf}.

Our next example is a relativistic theory of gauged fermions confined
to a defect.  Here a perturbative analysis near $d=2$ dimensions
reveals fixed point merger and BKT scaling.  A rainbow approximation
to the gap equation gives qualitatively similar results.

One of the motivations for this paper is to understand the chiral
phase transition that happens in (nonsupersymmetric) large-$N_c$ QCD
when the number of flavors of massless fermions $N_f$ varies.  As with
SQCD, we know there exists a conformal window for QCD in the parameter
$x=N_f/N_c$ where the upper end occurs at $x_*=11/2$, near
which the Banks-Zaks calculation is perturbative and reliable.  For
decreasing $x$ conformality must eventually be lost, since 
%the theory is asymptotically free 
for small $x$ chiral symmetry
breaking is expected. We speculate that the phase transition at this
lower boundary of the conformal window occurs due to fixed point
merger.  This suggestion is not new: it has been advocated before by
Gies and Jaeckel based on the results from the functional RG
approach~\cite{Gies:2005as}.
%\footnote{For an earlier speculation on the fate of the Banks-Zaks
%fixed point at strong coupling, see Ref.~\cite{Milton:1667}.}
If this picture is correct, then near
the transition the chiral condensate must exhibit BKT scaling.
Incidentally, this exponential behavior is also typically found when
one solves the gap equation obtained from (an unsystematic) truncation
of the Schwinger-Dyson hierarchy (see, e.g.,
\cite{Miransky:1984ef,Holdom:1988gs,Holdom:1988gr,Appelquist:1996dq,Appelquist:1998rb}, and
\cite{Miransky:2000rb} for further references).%
\footnote{In this context, BKT
scaling is sometimes called ``Miransky scaling.''}  A priori, the
relationship between the RG picture of merging fixed points and the
gap equation is not obvious; however our analysis of relativistic
defect fermions yields the same result in the regime where both
approaches are reliable.

If QCD does indeed exhibit BKT scaling, then our arguments suggest
that within the conformal window there exists another theory, QCD$^*$,
which is defined at the UV fixed point.  We conclude with speculations
about this theory.

\section{The BKT phase transition}
\label{sec:BKT}

The BKT phase transition~\cite{Berezinskii,Kosterlitz:1973xp} is due
to the deconfinement of vortices in the XY model at a critical
temperature $T_c$, above which the theory is conformal.  The behavior
of the correlation length \eq{BKT} below the phase transition can be
understood from the appropriate RG
equation~\cite{ZinnJustin:2000dr}.  We can exploit the equivalence
between the XY model and the zero temperature sine-Gordon model in
$1+1$ dimensions:
\beq
  L = \frac{ T}{2} (\d_\mu\phi)^2 - 2z \cos\phi\, ,
\eeq
where $T$ corresponds to the temperature of the XY model in units of
the spin coupling. Near the phase transition, it is useful to use the
variables $u=1-1/8\pi T$ and $v=2z/T\Lambda^2$ --- where $\Lambda$ is
the UV cutoff associated with the vortex core --- in terms of which
the perturbative $\beta$-functions are
\beq
\beta_u=-2v^2 ,
\qquad \beta_v=-2uv\, .
\eeq
Changing variables to $v+u=\tau$ and $v-u=2w$, one sees that $\tau w$
invariant under RG flow, and the running of $\tau$ is governed by
\beq\eqn{BKT-RG}
\beta(\tau; w\tau)= \mu \frac{d\tau}{d\mu} = -2 w\tau - \tau^2.
\eeq
This $\beta$-function has exactly the quadratic form of our toy model
\eq{toy}, with the substitution
\beq
(\alpha-\alpha_*) \to -2 w\tau , \quad (g-g_*) \to \tau.
\eeq
However, this $\beta$-function is only valid for small $\tau$ and $w$,
so the region about $\tau=0$ is excluded for fixed $w\tau$, as shown
in Fig.~\ref{fig:BKTbeta}.  Because of the excluded region, the
physics for the BKT model is slightly different than for the toy
model: in the non-conformal phase ($w\tau>0$), instead of starting
from the left of $\tau_*=0$ in the UV and flowing to the right in the
IR, the system starts at the top of the hill just to the right of
$\tau_*$ and flows to the right in the IR.  While it may appear that
this requires a fine-tuned initial condition for $\tau$, that is not
the case in terms of the $u$ and $v$ variables.  Starting the flow
near $\tau=0$ gives a factor of $1/2$ in the exponent for the
correlation length relative to the expression \eq{toycorr}:
%%%%%%%%
\begin{figure}[t]
\includegraphics[width=0.40\textwidth]{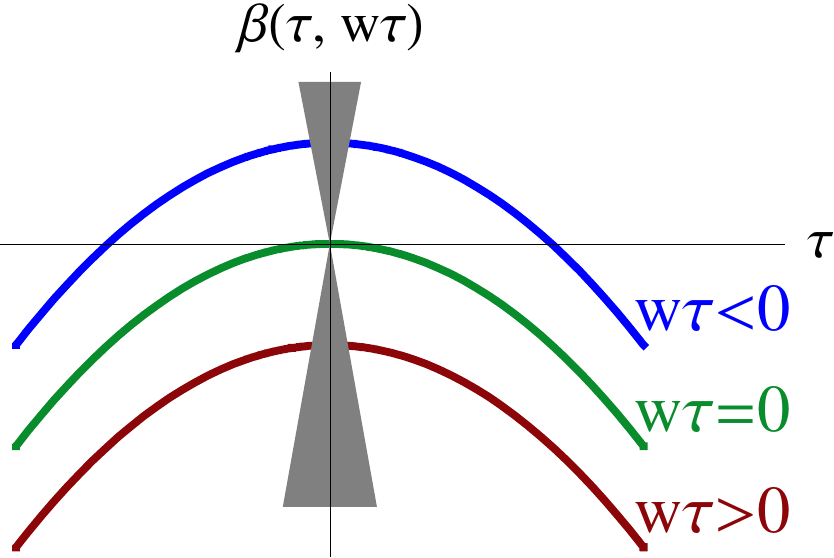}
\caption{The function $\beta(\tau)$ in the vicinity of $\tau=0$ for
the BKT transition \eq{BKT-RG}; the gray region is outside the realm
of validity of the calculation. }
\label{fig:BKTbeta}
\end{figure}
%%%%%%%%%%%
\beq
\xi_\text{BKT}\Lambda \simeq e^{\pi/(2\sqrt{2w\tau})}\ .
\eeq
The critical temperature is found by solving $w\tau=0$; expanding
about $T=T_c$ yields the familiar BKT result
\beq
\xi_\text{BKT}\Lambda \simeq e^{b'/|T-T_c|^{1/2}}\ , 
\eeq
where $b'$ is a nonuniversal number that can be expressed in terms of
$z$ and $\Lambda$.

\section{A nonrelativistic example: quantum mechanics in $1/r^2$ potential}
\label{sec:QM}

It is well known that the solutions for a quantum particle in a potential 
\beq
V(r)=\alpha/r^2
\eqn{vr2}
\eeq
possess conformal symmetry when the potential is repulsive or weakly
attractive $(\alpha > \alpha_*)$, but that for sufficiently attractive
potential $(\alpha<\alpha_*)$, conformality is lost and the potential
has discrete bound states.\footnote{There is a vast literature on the
$1/r^2$ potential.  For textbook treatment, see Ref.~\cite{LL3:1977};
for an early reference, see ~\cite{Case:1950an};
for relatively recent RG treatments see \cite{Beane:2000wh,
Barford:2002je, Bawin:2003dm,Barford:2004fz,Hammer:2008ra} and
references therein.}
%~\cite{essin2006qmx,coon2002aqm}.
% In $d=3$ spatial dimensions, for example, $\alpha_*=-\frac{1}{4}$. 
For a range of $\alpha$, the zero-energy, $s$-wave solution to
the Schr\"odinger equation for two particles with mass $m=1$ in
$d$ dimensions interacting via the potential $V(r)$ is given by
\beq
  \psi ={c_-}{r^{\nu_-}} +{c_+}{r^{\nu_+}}\, ,
  \qquad \nu_\pm =-\frac{( d-2)}{2}\pm\sqrt{\alpha-\alpha_*}\ ,\qquad 
 \alpha_* \equiv -\frac{(d-2)^2}{4}\ .
 \eqn{zesol}
  \eeq
  This solution is valid for $\alpha$ in the range
  \beq
  \alpha_*\le \alpha\le \alpha_*+1\ ;
 \eqn{arange}
  \eeq
for $\alpha <\alpha_*$ the above solution becomes complex, and the
Hamiltonian does not have ground state, while for $\alpha>\alpha_*+1$
then $\nu_- < -d/2$ and the $r^{\nu_-}$ solution is not normalizable
near $r\sim 0$.  Within the range \eq{arange}, if either $c_+$ or
$c_-$ vanish, then the solution is scale-invariant.  Solutions for
which both $c_+$ and $c_-$ are nonzero define an intrinsic length
scale, $L\equiv (c_+/c_-)^{1/(\nu_--\nu_+)}$ and therefore do not
exhibit conformal invariance; however in this case the solution always
approaches $c_+ r^{\nu_+}$ for large $r$ (since $\nu_+ \ge \nu_-$) and
so we can identify the $c_-=0$ solution with an IR attractive fixed
point and the $c_+=0$ solution with a UV fixed point, in a manner we
can make precise.  Arranging to have one of these solutions or the
other requires different boundary conditions at the origin, so we see
that the theory is actually not well defined with the potential
\eq{vr2}, but that it must be augmented by a $\delta$ function at the
origin which controls the boundary condition at $r=0$:
\beq
V(r)=\alpha/r^2 - g\delta^d(r)\,.
\eqn{vr2d}
\eeq
We will show that the coupling $g$ obeys an RG equation analogous to
our toy model \eq{toy}, and that the two conformal solutions $c_-=0$
and $c_+=0$ will correspond to two different fixed points of the
coupling $g$.  As $\alpha$ approaches $\alpha_*$ from above, we will
show that the two fixed points merge, $g_\pm=g_*$, at a value for
$g_*$ which we will compute.  For $\alpha<\alpha_*$ a UV cutoff must
be imposed on the theory in order to have a ground state and an IR
scale emerges which is related to the UV cutoff through the BKT
scaling formula \eq{toycorr}.  We show this in two different ways:
first we perform a nonperturbative analysis, and then we use Feynman
diagrams in a perturbative calculation in $2+\epsilon$ dimensions.
Both calculations shed light on the relativistic example we provide
later, and on our conjecture about the behavior of QCD as a function
of the number of flavors.

\subsection{A nonperturbative calculation}
\subsubsection{The exact wavefunction and energy}

To solve the Schr\"odinger equation exactly for two particle
scattering via a $1/r^2$ potential in $d$ dimensions we need to
regulate the singularity at $r=0$.  We choose to do so by considering
the potential
\beq
V(r) = \begin{cases}{\alpha/r^2}, & r>r_0,\cr -{g/r_0^2}, & r< r_0,\end{cases}
\eqn{qmpot}
\eeq
where $r_0^{-1}$ will serve as the cutoff $\Lambda_\text{UV}$.

At low energy, there is a region $r_0<r\ll 1/\sqrt{E}$ where the
$E\psi$ term in the Schr\"odinger equation can be neglected, and for
$\alpha>\alpha_* $ we find the solution \eq{zesol}
\beq
  \psi = {c_-}{r^{\nu_-}}+{c_+}{r^{\nu_+} }\, ,
\eeq
with the ratio $c_+/c_-$ given in terms of Bessel functions as
\beq
  \frac{c_+}{c_-} =-r_0^{(\nu_--\nu_+)} \frac{\gamma + \nu_-}{\gamma + \nu_+}
  \ ,\qquad
  \gamma \equiv\left[ \frac{\sqrt{g} J_{\frac{d}{2}}\left(\sqrt{g}\right)}
  {J_{\frac{d-2}{2}}\left(\sqrt{g}\right)}\right] .
%  
%  
%  
%  \frac{\sqrt{g} J_{\frac{d}{2}}
%  \left(\sqrt{g}\right)-\nu_- J_{\frac{d-2}{2}}\left(\sqrt{g}\right)}
%  {\sqrt{g} J_{\frac{d}{2}}\left(\sqrt{g}\right)-\nu_+ J_{\frac{d-2}{2}}
%  \left(\sqrt{g}\right)}
%   \equiv r_0^{(\nu_+-\nu_-)} F(g,\alpha,d)\, .
 % - \frac{\sqrt g \cot\sqrt g -1+\nu_-}{\sqrt g \cot\sqrt g -1+\nu_+}\,
  % r_0^{\nu_+-\nu_-} .
\eqn{gamdef}\eeq

The quantity $(c_+/c_-)$ is a dimensionful quantity
characterizing this solution; by requiring that it does not change as we
change the UV cutoff $r_0$, we arrive at the exact $\beta$-function for
$\gamma$ (defining RG time $t = -\ln r_0$):
\beq
\beta_\gamma= \frac{\d \gamma}{\d t} = -(\gamma+\nu_+)(\gamma+\nu_-) = 
\left(\alpha-\alpha_*\right) - (\gamma-\gamma_*)^2\ ,
\eqn{bdex}
\eeq
with 
\beq
\alpha_* = -\left(\frac{d-2}{2}\right)^2\ ,\qquad \gamma_* 
  = \frac{d-2}{2}\ ,\qquad \gamma_\pm = -\nu_\mp 
  = \frac{d-2}{2}\pm\sqrt{\alpha-\alpha_*}\ .
\label{alphastar}
\eeq
We recognize this to be the same $\beta$-function as our toy model
\eq{toy} with fixed points at $\gamma_\pm$; referring to \eq{gamdef}
we see that the IR fixed point corresponds to $\gamma=\gamma_-$ and
$c_-=0$, while the UV fixed point is associated with $\gamma=\gamma_+$
and $c_+=0$.

For general $d$ and $\alpha<\alpha_*$, scaling solutions do not exist;
physical quantities, such as the bound state energy, depend on the UV
cutoff.  Motivated by the discussion of coupling constant flow in our
toy model, we know that physical quantities will be insensitive to the
value of $\gamma$ (the UV coupling) so long as $\gamma< \gamma_*$, as seen in
Fig.~\ref{fig:gflow}.  So we take $\gamma\to -\infty$, which
is reached in the limit of a 
hardcore repulsive potential for $r< r_0$, $g\to-\infty$.
The ground state
wavefunction is then described by the Bessel function $\psi (r) =
r^{-(d-2)/2}K_{i\eta}(k r)$ with $\eta=\sqrt{\alpha_*-\alpha}$ 
and the boundary condition $\psi(r_0)=0$.  For
small real $\eta$ we can solve for $k$ and
find the binding energy
\beq
  B=k^2 = \frac1{r_0^2} 
  \exp\left(-\frac{2\pi}{\sqrt{\alpha_*-\alpha}}+O(1)\right),
\eeq 
Note that this scale for the binding energy is easily attained from
the RG analysis as
\beq
B \simeq \Lambda_\text{IR}^2=\left(\frac{1}{r_0} 
e^{\int_{-\infty}^{\infty} d\gamma/\beta_\gamma}\right)^2 
= \frac{1}{r_0^2} e^{-  2\pi/\sqrt{\alpha_*-\alpha}}\ .
\eeq
If one takes $\gamma$ to be arbitrarily close to $\gamma_*$ in the UV, 
and then takes $\alpha\to\alpha_*$ then the binding energy goes
to zero, but the exponent is only half as large,
\beq
  B \sim  \frac1{r_0^2} \exp\left(-\frac{\pi}{\sqrt{\alpha_*-\alpha}}\right).
\eeq
recalling the result  for the BKT transition.

\subsubsection{Onset of the Efimov effect}

Although the $\beta$-function in \eq{bdex} takes the same form as the
toy $\beta$-function in \eq{toy}, \eq{gamdef} implies that the
coupling $g$ is a multi-valued function of $\gamma$, with $|\gamma|\to
\infty$ identified with the zeros of
$J_{\frac{d-2}{2}}\left(\sqrt{g}\right)$.  Therefore for $\alpha <
\alpha_*$ our RG equation actually describes limit cycle behavior: as
$\gamma$ runs from $-\infty$ to $+\infty$ in RG period $T =- \int
d\gamma/\beta_\gamma = \pi/\sqrt{\alpha_*-\alpha}$, $g$ runs from one
Bessel function zero to the next.  It follows that there is not just
one IR scale defined by this RG flow, as in Fig.~\ref{fig:gflow}, but
an infinite number of such scales, each successively smaller than the
previous by a factor of $\text{exp}[-\pi/\sqrt{\alpha_*-\alpha}]$.
This behavior can explain the Efimov effect in 3-body bound states.

The classic Efimov effect~\cite{Efimov:1970zz} concerns the system of
three identical bosons.  When the scattering length between two bosons
becomes large, the three-body system develops a series of ever
shallower bound states. This occurs because the three particles
interact via an $\alpha/r^2$ potential for $r\gg r_0$, where $r_0$ is
the two-body effective range, and $\alpha$ a fixed number satisfying
$\alpha<\alpha_*$.  These systems require a 3-body interaction, and
the renormalization of this interaction exhibits the limit-cycle
behavior discussed above. The infinite tower of IR scales is
associated with the infinite number of ``Efimov states" below
threshold, exhibiting a geometric spectrum \cite{Hammer:2008ra}.  Such
states have been observed in systems of trapped atoms tuned to a
Feshbach resonance.

Three degenerate bosons tuned to infinite scattering length (so-called
``unitary bosons'') do not have a variable $\alpha$ parameter; in
order to see a transition very similar to what happens at
$\alpha=\alpha_*$ we need a case when the Efimov effect appears as one
changes a tunable parameter.  This is realized by nonrelativistic
fermions at unitarity with different masses for two spin components,
$M$ (heavy) and $m$ (light).  The Efimov effect occurs in the $p$-wave
channel for two heavy and one light fermions if
$M/m>13.6$~\cite{Efimov:1973}.

It is known that for $8.6<M/m<13.6$ one can additionally fine tune the
three-body interaction to resonance~\cite{Nishida:2007mr}.  From our
point of view, the two theories with and without fine-tuning in the
three-body channel correspond to the UV and IR fixed points.  When
$M/m\to13.6$, the two fixed points approach each other: the difference
between theories with and without 3-body fine-tuning becomes smaller
and smaller.  Finally when $M/m>13.6$ the fixed point completely
disappear, and an energy scale appears in the problem: the ground
state energy of the three-body bound state.

%This includes the case of a repulsive core, $g<0$.  When $g<g_*$, the
%ground state energy remains finite as $\alpha\to-1/4$.

%
%We finish by remarking that the analogy between the $\beta$-function
%of our toy model \eq{toy} and the exact $\beta$-function \eq{bex} can
%be made exact with a change of variables
%\beq
%\gamma\equiv 1-\sqrt{g}\cot\sqrt{g}\, .
%\eeq
%The $\beta$-function for $\gamma$ is
%\beq
%\beta_\gamma = \frac{\d \gamma}{\d t} = \left(\alpha+\frac{1}{4}\right)
%- \left(\gamma-\frac{1}{2}\right)^2,
%\eeq
%which has $\alpha_*=-1/4$ still, and $\gamma_*=1/2$.  However, note
%that $\eta$ is a a periodic function of $g$, and that as $\sqrt{g}$
%passes through $n \pi$, where $n$ is an integer, $\gamma$ goes from
%$+\infty$ to $-\infty$.  So in fact the RG flow of $\gamma$
%corresponds to a limit cycle of period
%\beq
%T=\int_{-\infty}^\infty \frac{d\gamma}{\beta_\gamma} 
%=\frac{\pi}{\sqrt{\alpha_*-\alpha}}\,.
%\eeq
%This periodic behavior is physical, and the same mechanism underlies
%the Efimov effect in a three-body system of particles with infinite
%2-body scattering lengths.

\subsubsection{Operator anomalous dimensions at the IR and UV fixed points.}

We can gain insight about the two fixed points by looking at the
dimension of the operators.  Let us consider the two-particle operator
$\psi\psi$.  According to the operator/state correspondence
developed in Ref.~\cite{Nishida:2007pj}, one can find dimensions of
this operator by putting two particles in a harmonic potential.
%For convenience we choose $\alpha=\ell(\ell+1)$ used in previous section. 
%Fixed points are located at
%\beq
%\sqrt{g_*}\cot \sqrt{g_*}=-\ell ~~~~~\textrm{and}~~~~~ 
%\sqrt{g_*}\cot \sqrt{g_*}=\ell +1.
%\eeq
The Hamiltonian of the system is given by
\beq
H=-\frac{1}{2}{\bf\nabla}^2_1-\frac{1}{2}{\bf\nabla}^2_2
+V(|{\bf r}_1-{\bf r}_2|)+\frac{1}{2}\omega^2(r_1^2+r_2^2)\, .
\eeq
%where the mass of the particle has been absorbed to the Hamiltonian.
In terms of the center of mass coordinate ${\bf R}$ and relative
coordinate ${\bf r}$, the Hamiltonian can be rewritten as $H=H_R +
H_r$ where the ground state energy of $H_R$ equals $d\omega/2$ and
\beq
H_r=-{\bf\nabla}^2_r+V(r)+\frac{1}{4}\omega^2r^2,
\eeq
where the potential is given in \eq{qmpot}.  The ground state
wavefunctions and energies for this Hamiltonian for $g$ tuned to one
of the fixed points $g_\pm$ is easily seen to equal
\beq
\psi_\pm = {e^{-\omega r^2/4}}{r^{\nu_\pm}}\ ,\qquad 
  E_r^\pm = \left(\frac{d}{2}+\nu_\pm\right)\omega
\eeq
in the limit $r_0\to 0$. (Recall that the fixed points $g_\pm$
correspond to solutions $\psi=r^{\nu_\pm}$ in the absence of the
harmonic potential).
Therefore the total ground state energy is
\beq
E^\pm=\left(d+\nu_\pm\right)\omega\,,
\eeq
and so the scaling dimensions of the two-particle operator $\psi\psi$ are
\beq
\Delta_{\pm}= \left(d+\nu_\pm\right) 
  = \frac{d+2}{2} \pm \sqrt{\alpha-\alpha_*}\, ,
\eeq
where $\Delta_+$ and $\Delta_-$ are the operator dimension at the IR
and UV fixed points respectively.  We emphasize the fact that
\beq
(\Delta_++\Delta_-)=d+2
\eeq
in any spatial dimension $d$ and for any in the range $ \alpha_*\le
\alpha<( \alpha_*+1)$ in \eq{arange}, with $\Delta_+=\Delta_-=(d+2)/2$
at $\alpha=\alpha_*$.  Note that $(d+2)$ is the scaling dimension of a
nonrelativistic Lagrange density, since time has twice the scaling
dimension as space; we return to this below, when we discuss the
AdS/CFT correspondence.

%Fig.~\ref{fig:scalingdimension}.
%\begin{figure}
%\includegraphics[width=0.5\textwidth]{scalingdimension}
%\caption{Plot of scaling dimension of $\psi\psi$ with respect to $\alpha$. 
%The upper and lower lines represent $\Delta_+$(UV fixed) and 
%$\Delta_-$(IR fixed), respectively.}
%\label{fig:scalingdimension}
%\end{figure}
%
%(describe the calculation of $\Delta_+$ and $\Delta_-$, with special
%mention on the fact that $\Delta_++\Delta_-$ is constant and at
%$\alpha=-1/4$, $\Delta_+=\Delta_-$). 

%(a plot of dimensions as function of $\alpha$).
%

%\subsubsection{Range of $\alpha$ where the fine-tuned fixed point exists}

%We found that for $\alpha>\alpha_*$ there are two scale-free boundary
%conditions one can impose on the wave functions near the origin.  We
%identified the less singular behavior $1/r^{\nu_-}$ with to the
%IR fixed point.  It turns out that the identification of the more
%singular behavior, $1/r^{\nu_+}$, with the UV fixed point is valid
%only for a range of $\alpha$.  This comes from the requirement that
%the wave function $\psi(r)$ does not tend to zero when one takes
%$r_0\to0$.  This happens only when the normalization integral
%$\int\!d{\bf r}|\psi({\bf r})|^2$ is dominated by the tail, but not by
%the core $r<r_0$.  This requirement translates into
%\begin{equation}
%  \nu_+ < \frac d2\,,
%\end{equation}
%or
%\begin{equation}
%  \alpha_* < \alpha < \alpha_*+1, \qquad \alpha_* = -\frac{(d-2)^2}4\,.
%\end{equation}
%Only in this range of $\alpha$, there is a UV fixed point.  For
%$\alpha>\alpha_*>1$, only the IR fixed point exists.  

%

\subsection{The renormalization group: $\epsilon$ expansion}

When we consider relativistic quantum field theories a nonperturbative
solution will not be available, and we will have to rely on either
perturbation theory, or a truncation of the Schwinger-Dyson equations.
It is therefore instructive to examine a perturbative analysis of the
$1/r^2$ potential.  We have seen that the doubly degenerate fixed
point at the phase transition occurs at coupling $\gamma=\gamma_* =
(d-2)/2$; we therefore start with the action in $d=2+\epsilon$ spatial
dimensions, where perturbation theory can correctly describe the phase
transition.  In order to facilitate the use of Feynman diagrams, we
write the theory in second quantized form with a contact interaction,
\begin{multline}
  S = \int\!dt\,d^d\x\, \left( i\psi^\+\d_t\psi - \frac{|\nabla\psi|^2}{2}
      + \pi\frac g{4} \psi^\+ \psi^\+ \psi \psi
     \right) \\
     - \int\!dt\,d^d\x\,d^d\y \psi^\+(t,\x)\psi^\+(t,\y) 
     \frac\alpha{|\x-\y|^2} \psi(t,\y)\psi(t,\x),
\end{multline}
where the factor of $\pi$ in the contact interaction is chosen for
future convenience.  The Feynman rules are as follows:
\begin{itemize}
\item Propagator 
\beq
  \frac i{\omega - \p^2/2}\,,
\eeq
\item Contact vertex
\beq
  i \pi g\mu^{-\epsilon},
\eeq
\item ``Meson exchange''
\beq
  \frac{2\pi i\alpha}\epsilon \frac1{|\q|^\epsilon}\,.
\eeq
\end{itemize}
Note the unusual $1/\epsilon$ pole in the ``meson propagator.''  It
arises because the Fourier transform of a $1/r^2$ interaction is log
divergent in $d=2$ dimensions.

\begin{figure}[t]
\includegraphics[width=0.45\textwidth]{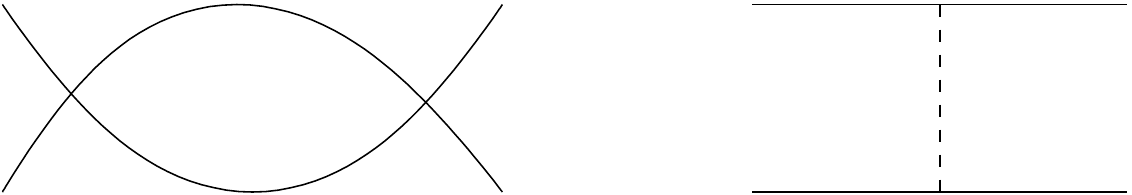}
\caption{Two diagrams contributing to the $\beta$-function in
Eq.~(\ref{eq:beta1}).  Note that the second diagram is a tree diagram.}
\label{fig:beta1}
\end{figure}

It is easy to see that $\alpha$ does not get renormalized (as one
would expect, being the strength of a nonlocal interaction); however
the coupling $g$ runs.  From the above Feynman rules, the
$\beta$-function for $g$ arises from the sum of the tree graph and the
one-loop graph shown in Fig.~\ref{fig:beta1}, with the result
\beq
   \beta(g;\alpha) =\frac{\d g}{\d t} =  \epsilon g - \frac{g^2}{2} + 2\alpha
  = 2\left(\alpha+\frac{\epsilon^2}{4}\right)-\frac{1}{2}(g-\epsilon)^2 \ ,
\eqn{beta1}\eeq
which we recognize to be equivalent to our toy model, up to an
unimportant rescaling of $g$ by $2$, with
\beq
g_* = \epsilon\ ,\qquad \alpha_*= -\frac{\epsilon^2}4\ ,
\eqn{crit}
\eeq
%in agreement with \eq{astar}. 
Note that our perturbative expansion is
justified for small $\epsilon$, but $\alpha_*$ coincides with the
exact result in Eq.~(\ref{alphastar}).
For $\alpha>\alpha_*$ the
$\beta$-function has two zeros: $g_\pm =g_*\pm
2\sqrt{\alpha-\alpha_*}$.  At $\alpha=0$, $g_-=0$ is the IR stable
fixed point, corresponding to a noninteracting theory --- for the
generic short-ranged potential, low-energy scattering is trivial;
$g_+=2\epsilon$ corresponds to a fine-tuned potential with a bound
state at threshold (e.g., an infinite scattering length).  As one
decreases $\alpha$ the two fixed points approach each other, merging
at $g_\pm=g_*$ when $\alpha=\alpha_*$.

For $\alpha<\alpha_*$ the potential requires a cutoff and has a bound
state; we can estimate the size of the bound state to be given by the
correlation length $\xi=\Lambda_\text{IR}^{-1}$ in \eq{toycorr}; this
gives a binding energy $B\sim \Lambda_\text{IR}^{2}$ , or
\beq
  B \sim  {\Lambda_{\rm UV}^2}
  \exp\left( -\frac{2\pi}{\sqrt{\alpha_*-\alpha}}\right).
\eqn{Bpert}
\eeq
Note that this formula is independent of $\epsilon$ and therefore
appears to be independent of dimension.  In fact the above estimate is
verified in the nonperturbative calculation of the previous
section.

\begin{figure}
\includegraphics[width=0.875\textwidth]{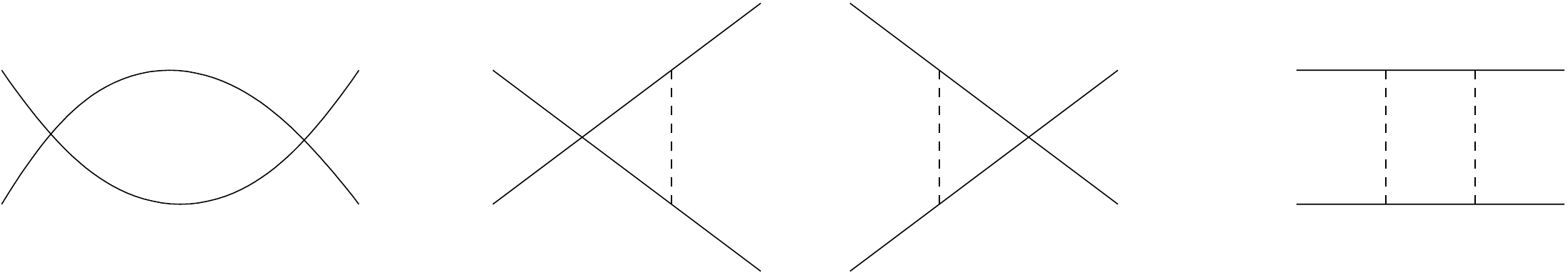}
\caption{Diagrams contributing to the perturbative $\beta$-function
for $\tilde g$.}
\label{fig:beta2}
\end{figure}

An unusual feature of our calculation of the $\beta$-function
(Fig.~\ref{fig:beta1}) is the contribution from a tree graph.  We
close this section by noting that a more conventional calculation is
obtained by making the following change of variable:
\beq
  g = \tilde g - \frac{2\alpha}\epsilon\,.
\eeq
Then the RG equation becomes
\beq
   \frac{\d\tilde g}{\d t} = \epsilon\tilde g 
  - \frac{{\tilde g}^2}{2} 
+2 {\tilde g} \left(\frac{\alpha}\epsilon\right)
  - \frac1{2} \left( \frac{2\alpha}\epsilon\right)^2.
\eeq
The first term on the the right hand side comes from the engineering
dimension.  The other terms come from the diagrams as in
Fig.~\ref{fig:beta2}.
All diagrams now have loops.  This is a more natural approach from the
point of view of Wilsonian RG, where one looks at the logarithm in the
momentum integral instead of the $1/\epsilon$ poles.  But we emphasize
that the two RG equations lead to the same physical consequences.

\subsubsection{Summary of the QM example}

Before proceeding, we summarize our findings.
\begin{itemize}
\item The theory has two fixed points, IR and UV, when 
 $\alpha>\alpha_* =-(d-2)^2/4$.
\item When there are two fixed points, the dimensions of the scalar
operators at the IR and UV fixed points are $\Delta_+$ and $\Delta_-$, 
and they satisfy $\Delta_++\Delta_-=d+2$.
\item When $\alpha<-1/4$, the fixed points do not exist and  the theory
develops a bound state energy which scales as 
  $\Lambda_\text{UV}^2\, \exp(-2\pi/\sqrt{\alpha_*-\alpha})$.
\end{itemize}

\section{A holographic perspective}
\label{sec:holo}

As far as we know, there is no simple holographic dual description of
quantum mechanics with $1/r^2$ potentials, nor of the field
theoretical models considered later in this paper.  However,
holography provides an interpretation of conformality loss which turns
out to be very useful in developing our intuition about such phase
transitions: the loss of conformality can be associated with the
violation of the Breitenlohner-Freedman (BF) bound.

\subsection{The conformal phase: pair of theories}

In our RG discussion, for $\alpha>\alpha_*$ there are two CFTs that
merge into one at $\alpha=\alpha_*$.  This situation is reminiscent of 
what occurs in holography~\cite{Klebanov:1999tb}: 
a higher dimensional theory containing a scalar field
$\phi$ with mass $m^2$ in the interval $-d^2/4<m^2<-d^2/4+1$
corresponds to \emph{two} different boundary theories in which the
dimensions of the operator $O$ dual to $\phi$ have two different
values
\begin{equation}\label{Deltam}
  \Delta_\pm = \frac d2 \pm \frac12 \sqrt{d^2+4m^2} \equiv 
  \frac d2 \pm \nu\,.
\end{equation}
The main point here is that in the asymptotics of the scalar field
near the boundary $z=0$ of the AdS space, $\phi(z)=c_- z^{\Delta_-}+
c_+ z^{\Delta_+}$, one can interpret $c_-$ as the source coupled to
$O$, and $c_+$ as its expectation value, and vice versa.

Instead of repeating the discussion in Ref.~\cite{Klebanov:1999tb}, we
illustrate its main points in a simple model.
% where the analogy with
%quantum mechanics in $1/r^2$ potential is quite explicit.  
In this
model, one sees that the theory with $[O]=\Delta_-$ can be
obtained from the theory with $[O]=\Delta_+$ by adding to the
Lagrangian a term $O^2$ with a fine-tuned coefficient (in other words,
we will go ``against the RG flow,'' cf.\ Ref.~\cite{Gubser:2002vv}
where one follows the RG flow from the UV fixed point to the IR fixed
point).

Consider a massive scalar field in AdS$_{d+1}$ space.  We use Euclidean
signature in this subsection, so the metric is
\begin{equation}
  ds^2 = \frac{R^2}{z^2}(dz^2 + dx^\mu dx^\mu)\,.
\end{equation}
We will set the radius of the AdS space $R=1$.  The action for the 
scalar field $\phi=\phi(z,x)$ is
\begin{equation}
\begin{split}
  S & = \frac12 \int\!dz\, d^dx\, \sqrt{g}( g^{\mu\nu}\d_\mu\phi\d_\nu\phi
      + m^2\phi^2) - \frac1{\epsilon^{\Delta_+}}
      \int\!d^dx\, J(x) \phi(\epsilon,x)\\
    & = \frac12 \int\!dz\,d^dx\, \frac1{z^{d+1}} \left[z^2(\d_z\phi)^2
      + z^2(\d_\mu\phi)^2 + {m^2}\phi^2\right]
      - \frac1{\epsilon^{\Delta_+}}\int\!d^dx\, J(x) \phi(\epsilon,x)\,.
\end{split}
\end{equation}
In our model, this action is taken as the the definition of the CFT.
This CFT ``lives'' on the boundary in the sense the external source
$J$ couples only to the field at some small $z=\epsilon$, with
$1/\epsilon$ playing the role of the momentum UV cutoff.  The operator
$O(x)$ that $J$ couples to is defined as
$O(x)=\epsilon^{-\Delta_+}\phi(\epsilon,x)$.  The extra power of
$\epsilon$ is chosen so that subsequent results have a regular
$\epsilon\to0$ limit.

We assume a large $N$ parameter so that one can use the saddle point
approximation, in which $\phi$ satisfies the field equation
\begin{equation}
  \phi'' - \frac{d-1}z \phi' - q^2 \phi - \frac{m^2}{z^2}\phi 
  + \epsilon^{\Delta_- -1}J\delta(z-\epsilon) = 0\,,
\end{equation}
where we have changed to momentum space.  We assume $q\epsilon\ll1$,
i.e., $q$ is much smaller than the UV cutoff.  To completely specify
the solution we impose two boundary conditions.  Near $z=0$ there are
two possible solutions to this equation, $\phi\sim z^{\Delta_\pm}$
where $\Delta_\pm$ are defined in Eq.~(\ref{Deltam}).  We require that
\begin{equation}\label{bc-UV}
  \phi =c_0 z^{\Delta_+}, \qquad z\to 0\,,
\end{equation}
i.e., we require $\phi$ to follow the most regular asymptotic
behavior at small $z$.
We leave the boundary condition at $z\to\infty$ for later discussion.
Equation~(\ref{bc-UV}) is valid for $z<\epsilon$, but due to the insertion
of a source at $z=\epsilon$, $\phi$ contains both asymptotics once $z$ is
larger than $\epsilon$,
\begin{equation}\label{phic+c-}
  \phi = c_+ z^{\Delta_+} + c_- z^{\Delta_-}, \quad  \epsilon<z\ll q^{-1}\,.
\end{equation}
Clearly, $c_-$ is proportional to the source $J$.  Matching
boundary conditions one finds
\begin{equation}\label{c-J}
  c_- = \frac J{\Delta_+-\Delta_-}\, .
\end{equation}
From the point of view of the interior region $z>\epsilon$,
Eqs.~(\ref{phic+c-}) and (\ref{c-J}) effectively fix the boundary
condition near $z=\epsilon$.  Here we obtain a key ingredient of the
AdS/CFT prescription: the coefficient in front of the $z^{\Delta_-}$
part of the field is the source coupled the the operator $\phi$.

The coefficients $c_0$ in Eq.~(\ref{bc-UV}) and $c_+$ in
Eq.~(\ref{phic+c-}) can be determined only after the boundary condition
at $z=\infty$ is fixed.  We can relate the expectation
value of $O$ with $c_+$:
\begin{equation}
  \< O \>|_J = \frac{\phi(\epsilon)}{\epsilon^{\Delta_+}}
  = \frac J{(\Delta_+-\Delta_-)\epsilon^{\Delta_+-\Delta_-}}
    + c_+ \,.
\end{equation}
Therefore, up to a singular contribution, the expectation value of $O$
is related to $c_+$.  The two-point function $\<OO\>$ is then
\begin{equation}
  \< OO\> = \frac{\d}{\d J} \< O \>|_J =
    \frac J{2\nu \epsilon^{2\nu}} + \frac{\d c_+} {\d J} \,.
\end{equation}

Let us now impose the boundary condition at $z\to\infty$.  To ensure finiteness
of the action, it is sufficient to require
\begin{equation}
  \phi(z,x)\to 0\,, \qquad z\to \infty\,.
\end{equation}
The saddle point solution is now completely determined
\begin{equation}\label{solutionD}
  \phi(z) = \begin{cases} 
    D \epsilon^{-\nu} K_\nu(q\epsilon) z^{d/2} I_{\nu}(qz), & z<\epsilon\,,\\
    D \epsilon^{-\nu} I_\nu(q\epsilon) z^{d/2} K_{\nu}(qz), & z>\epsilon\,,
\end{cases}
\end{equation}
with $D=J\epsilon^{-\nu}$.  This solution corresponds to
\begin{equation}
  c_- = \frac{J}{2\nu}, \qquad 
  c_+ = -\frac{\Gamma(1-\nu)}{\nu^2\Gamma(\nu)2^{1+2\nu}} J q^{2\nu}.
\end{equation}
The two-point function $\<OO\>$ is proportional to $q^{2\nu}$,
consistent with dimension of $O$ being $\Delta_+=d/2+\nu$.

Now we turn on a deformation $O^2$ with a coefficient that will be
fine-tuned to get another conformal field theory.  The action is now
\begin{equation}
  S = \frac12\int\!dz\,d^dx\, \frac1{z^{d+1}}\left[z^2(\d_z\phi)^2
      + z^2(\d_\mu\phi)^2 + {m^2}\phi^2 \right]
      -\int\!d^dx\,\left[\frac \lambda{2\epsilon^d} 
      \phi^2(\epsilon)+J\frac{\phi(\epsilon)}{\epsilon^{\Delta_-}} \right].
\end{equation}
Let us first set $J=0$.  The field equation is
\begin{equation}\label{finetuned-eq}
  -\phi'' + \frac{d-1}z \phi' + \frac{m^2}{z^2} \phi +q^2\phi
  -\frac \lambda\epsilon\delta(z-\epsilon)\phi = 0\,.
\end{equation}
One can integrate this equation from $z=0$ to larger $z$.  For
$z<\epsilon$, $\phi$ is purely $z^{\Delta_+}$, and for $z>\epsilon$ it
becomes a mixture of $z^{\Delta_+}$ and $z^{\Delta_-}$, the relative
weight of which depends on $\lambda$.  The most interesting value of
$\lambda$ is when $\phi$ is purely $z^{\Delta_-}$ for $z>\epsilon$.
This happens when $\lambda$ is fine-tuned to the critical value
\begin{equation}\label{lambda-crit}
  \lambda = \Delta_+ - \Delta_-\,.
\end{equation}
There is a quantum-mechanical interpretation of this fine-tuning.  If
one identifies $z$ as the radial coordinate $r$ of a two-dimensional space,
then Eq.~(\ref{finetuned-eq}) is the radial Schr\"odinger equation for
the wave function  $\psi=z^{-d/2}\phi$ of
a particle moving in a potential which is a sum of a $1/r^2$ piece
and a delta-shell piece,
\begin{equation}
  V(r) = \frac{\nu^2}{r^2} - \frac\lambda{\epsilon}\delta(r-\epsilon)\,.
\end{equation}
with $-q^2$ playing the role of the energy. 
The value~(\ref{lambda-crit}) corresponds to the case when the
potential has a zero-energy bound state.

Let $\lambda$ be fine-tuned to this value, and turn on the source $J$.
Using the asymptotics $\phi\sim z^{\Delta_+}$ for $z<\epsilon$ and
integrating the field equations passed $z=\epsilon$, we find that for
$z>\epsilon$, the coefficient $c_+$ is now proportional to $J$:
\begin{equation}
  c_+ = -\frac {J}{\Delta_+-\Delta_-}\, .
\end{equation}
The expectation value for $O$ is now related to $c_-$,
\begin{equation}
  \< O \>_J = c_- \, . 
\end{equation}
The assignment of source and expectation value is reverse
to the case $\lambda=0$.
If one imposes the boundary condition $\phi(z)\to0$ when $z\to\infty$,
then the solution to Eq.~(\ref{finetuned-eq}) is given by
Eq.~(\ref{solutionD}), but now 
\begin{equation}
   D = \frac{J\epsilon^\nu}{1-2\nu I_\nu(q\epsilon) K_\nu(q\epsilon)}\,.
\end{equation}
The solution corresponds to
\begin{equation}
  c_+ = -\frac{J}{2\nu}\,, \qquad
  c_- = \frac{2^{2\nu-1}\Gamma(\nu)}{\Gamma(1-\nu)}\frac J{q^{2\nu}} \,.
\end{equation}
In particular $\<OO\>\sim q^{-2\nu}$, corresponding to
$[O]=\Delta_-=d/2-\nu$.

Thus, in this simple holographic model, the UV stable fixed point of
the CFT with the fine-tuned $O^2$ interaction corresponds to the same
bulk theory, but with the opposite assignment for the source and the
expectation value.

\subsection{Below the Breitenlohner-Freedman bound}

Here we speculate on the fate of the bulk theory with a scalar with
$m^2$ below the Breitenlohner-Freedman (BF) bound $-d^2/4$.  The most
interesting case is when $m^2$ is only slightly below the BF bound,
where the boundary theory is approximately conformal over a large
energy range.  The dual bulk description should involve a spacetime
that is approximately AdS, cutoff both at the UV and the IR by the
respective ``walls.''

First, for the set up with a scalar $m^2$ below the BF bound, there
must be an UV cutoff in the theory.  For example, the theory with a
dual description can arise as a low-energy limit of another theory
whose UV is free of any instability.  Let us model that by imposing a
hard cutoff on the AdS space, and impose a boundary condition on the
scalar $\phi$ at the cutoff.  The precise form of the boundary
condition is not important, for definiteness we take it to be
Dirichlet: $\phi(z_{\rm UV})=0$.

One expect that a IR scale will be generated by the condensation of
$\phi$.  We model that scale very roughly by another, IR, cutoff at
$z_{\rm IR}$, and impose another Dirichlet boundary condition there.

Now let us look at the field equation for $\phi$,
\beq
  \phi'' - \frac{d-1}z \phi' - \frac{m^2}{z^2}\phi-q^2 \phi = 0\,.
\eeq
Changing variables to $\phi=z^{(d-1)/2}\psi$, this equation becomes
\begin{equation}
  -\psi'' + \frac{m^2 + (d^2 - 1)/4}{z^2} \psi
  = - q^2 \psi\,.
\end{equation}
This equation, with the boundary condition on at $z_{\rm IR}$ and
$z_{\rm UV}$, gives us an infinite tower of particles.  The mass
square of the particles in this tower is the eigenstate of a particle
in a one-dimensional potential which is $\alpha/r^2$ enclosed between
two infinite walls at $z_{\rm UV}$ and $z_{\rm IR}$.  The condition of
absence of tachyon is equivalent to the condition that the potential
does not contains a negative-energy eigenstate.  This requires the
interval between the two cutoff is not too large,
\beq
   \ln\frac{z_{\rm IR}}{z_{\rm UV}} < \frac{\pi}{\sqrt{m_{\rm BF}^2-m^2}}\,.
\eeq
In a more realistic setup where the scale $z_{\rm IR}$ appears
dynamically, one can expect that it appears at the scale requires for
preventing a tachyon,
\begin{equation}
  z_{\rm IR} \sim z_{\rm UV}\exp\biggl( 
  \frac{\pi}{\sqrt{m_{\rm BF}^2-m^2}}\biggr).
\end{equation}

We expect that this situation is rather generic in holography.  It
would be interesting to construct an explicit solution in string
theory which exhibits the BKT scaling.

\section{A relativistic example: defect QFT}
\label{sec:defect}

In this section we consider a relativistic quantum field theory that
exhibits the phenomenon of fixed point annihilation.  The example
resembles QCD with large number of flavors, but the phase transition
occurs in the regime of weak coupling.

We consider a theory of a fermion living on a $d$-dimensional
membrane, and interacting through a SU($N_c$) gauge field that lives
in (3+1) dimensions.  We shall assume that the SU($N_c$) gauge
coupling does not run; it can be easily accomplished by taking the
gauge field to be part of a conformal field theory, say, the ${\cal
N}=4$ super-Yang-Mills theory.  The most interesting case is $d=3$
($=2+1$), which was analyzed by Rey using the Schwinger-Dyson
approximation and gauge/gravity duality~\cite{Rey}.  Here we shall
take $d=2+\epsilon$ (i.e., $(1+\epsilon)+1$) to take advantage of a
small parameter $\epsilon\ll1$.

The action is
\beq
  S = \int\!d^dx\, (i\bar\psi\gamma^\mu \d_\mu\psi +
      \bar\psi\gamma^\mu\psi A_\mu) 
  - \frac14 \int\!d^4x\, F_{\mu\nu}^a F_{\mu\nu}^a + \cdots,
\eeq
and we assume there is a UV cutoff $\Lambda$.  The $d$-dimensional
photon propagator is obtained by integrating out the 4d propagator
over transverse directions,
\beq\label{boundary-photon}
  D_{\mu\nu}(q) = \int\!\frac{d^{2-\epsilon}q_\perp}{(2\pi)^{2-\epsilon}}
        \frac{-i}{q^2}\left(g_{\mu\nu} - 
        (1-\xi)\frac{q_\mu q_\nu}{q^2}\right),
\eeq
where $q^2=q_\parallel^2 -q_\perp^2$.  For small $\epsilon$, the result is
\beq\label{Dmunu-eps}
  D_{\mu\nu}(q) = \frac{ig_{\mu\nu}}{2\pi\epsilon}
  \left( \frac1{(-q^2)^{\epsilon/2}} - \frac1{\Lambda^\epsilon}\right).
\eeq
Note that the dependence on the gauge parameter $\xi$ disappears in
the small $\epsilon$ limit (if one assumes $\xi\sim1$).

\subsection{Schwinger-Dyson treatment (rainbow approximation)}

\begin{figure}[b]
\includegraphics[width=0.28\textwidth]{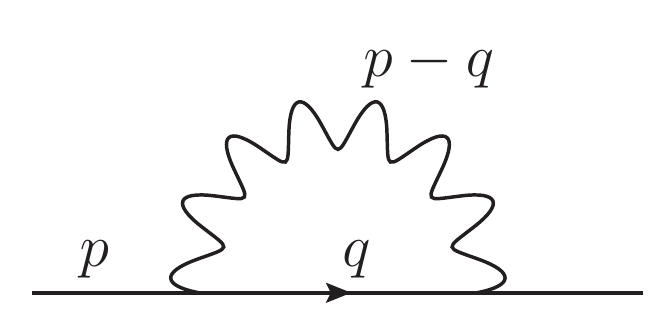}
\caption{The one-loop graph that contributes to the gap equation}
\label{fig:SD}
\end{figure}

Before going to the RG treatment, we review how the chiral phase
transition is found using the gap equation.  This treatment is very
similar to that used in QCD~\cite{Miransky:1984ef,Appelquist:1996dq}.
The lowest-order gap equation is (Fig.~\ref{fig:SD})
\beq
  -i \Sigma(p) = - \int\!\frac{d^dq}{(2\pi)^d}\, D_{\mu\nu}(p-q)
   \gamma^\mu t^a
  \frac{\gamma^\nu q_\nu + \Sigma(q)}{q^2 - \Sigma^2(q)+i\epsilon}
   \gamma^\nu t^a.
\eeq
Inserting the photon propagator~(\ref{boundary-photon}) and performing
a Wick rotation, the equation becomes, for small $\epsilon$,
\beq\label{gap-eq}
  \Sigma(p) = \frac{g^2 C_A}{4\pi^3\epsilon}\int\!d^dq\,
  \frac1{|p-q|^\epsilon} \frac{\Sigma(q)}{q^2+\Sigma^2(q)}\,,
  \qquad C_A \equiv \frac{N_c^2-1}{2N_c}\,,
\eeq
where the integral is taken in Euclidean space.  It will become clear
later that the dominant contribution to the integral comes from the
regions $p\ll q$ and $p\gg q$, with $p\sim q$ giving a subleading
contribution.  Changing variables to
\beq
  x = \ln\frac pm\,, \qquad y = \ln\frac qm\,,
\eeq
where $m=\Sigma(0)$ will be the mass gap, the Schwinger-Dyson equation
becomes
\beq\label{SD-int}
  \Sigma(x) = \frac{g^2C_A}{2\pi^2\epsilon} \left[
    \int_0^x\!dy\, [e^{-\epsilon(x-y)}-e^{-\epsilon(x_m-y)}]\Sigma(y)
  + \int_x^{x_m}\!dy\, 
  [1-e^{-\epsilon(x_m-y)}]
  \Sigma(y) \right],
\eeq
where $x_m=\ln(\Lambda/m)$ and $\Lambda$ is the UV cutoff.
Differentiating Eq.~(\ref{SD-int}) over $x$, we find
\begin{align}
  \Sigma'(x) & = -\frac{g^2C_A}{2\pi^2}\int_0^x\!dy\,
  e^{-\epsilon(x-y)}\Sigma(y)\,, \\
  \Sigma''(x) &= \frac{g^2C_A\epsilon}{2\pi^2} \int_0^x\!dy\,
  e^{-\epsilon(x-y)}\Sigma(y) - \frac{g^2C_A}{2\pi^2}\Sigma(x)\,,
\end{align}
from which we find that $\Sigma$ satisfies the differential equation
\beq
  \Sigma''(x) + \epsilon\Sigma'(x) + \frac{g^2C_A}{2\pi^2}\Sigma(x) = 0\,,
\eeq
with boundary conditions
\beq
  \Sigma'(0) = 0\,, \qquad
  \Sigma(x_m) = 0\,.
\eeq

The solution to the equation is
\beq
  \Sigma(x) = m e^{-\epsilon x/2} \frac{\cos (\kappa x-\delta)}{\cos\delta}\,,
\eeq
with
\beq
  \kappa = \sqrt{\frac{g^2C_A}{2\pi^2}-\frac{\epsilon^2}4}\,.
\eeq
When $\epsilon$ and $\kappa$ are small, $\Sigma$ varies slowly on the
logarithmic scale, which validates the assumption that the integral in
Eq.~(\ref{gap-eq}) is dominated by regions where $p$ and $q$ are very
different.

The boundary conditions imply
\beq
  \tan\delta  =  \frac\epsilon{2\kappa}\,,\qquad
  \cos(\kappa x_m -\delta) = 0\,,
\eeq
from which one finds
\beq
  x_m = \frac1\kappa\left[ \left( n+\frac12\right) 
    + \arctan\frac\epsilon{2\kappa} \right],
\eeq
where $n$ is an integer.  The solution with $n=0$ corresponds to the
biggest gap and is favored energetically.  The dynamically generated
mass gap is
\beq
  m \sim \Lambda \exp \left[ -\frac1\kappa \left( \frac\pi2 +
   \arctan\frac\epsilon{2\kappa}\right) \right], \qquad
  \kappa = \sqrt{\frac{g^2C_A}{2\pi^2}-\frac{\epsilon^2}4}\,.
\eeq
So we find that there is a phase transition occurring at
\beq
  g_*^2 = \frac{\pi^2\epsilon^2}{2C_A} \,,
\eeq 
and the critical behavior of the gap near $g=g_*$ conforms with 
BKT scaling.

\subsection{RG treatment: beyond the rainbow}

The RG equation can be written in a way very similar to the RG
equation for the QM example with $1/r^2$ potential.  One introduces an
extra four-fermi interaction into the Lagrangian
\beq
  S = \int\!d^dx\, \Bigl(i\bar\psi\gamma^\mu \d_\mu\psi +
      \bar\psi\gamma^\mu\psi A_\mu 
    - \frac c2 (\bar\psi \gamma^\mu t^a\psi)^2\Bigr) 
  - \frac14 \int\!d^4x\, F_{\mu\nu}^a F_{\mu\nu}^a + \cdots\,.
\eeq
The tree level one-gluon exchange contains a $1/\epsilon$ factor from
the gluon propagator~(\ref{Dmunu-eps}) and contributes to the beta
function for $c$:
\beq
  \beta(c) = \epsilon c - \frac{N_c}{2\pi}c^2 - \frac{g^2}{2\pi}\,.
\eeq
The phase transition occurs at $g=g_*$ where $\beta(c)$ has a double zero,
\beq
  g_*^2 =  \frac{\pi^2\epsilon^2}{N_c}\,.
\eeq
When $g>g_*$, we need to solve the RG equation,
\beq
  \frac{\d c}{\d\ln\mu} = \beta(c)\,,
\eeq
with the boundary condition that the bare four-fermi coupling is zero 
at the UV cutoff, $g(\Lambda)=0$.  The solution is
\beq
  c(\mu) = \frac{\pi\epsilon}N + \frac{2\pi}N \kappa \tan \left[
     \kappa\ln\frac\Lambda\mu - \delta \right],\qquad
  \kappa = \sqrt{\frac{g^2N_c}{4\pi^2}-\frac{\epsilon^2}4}\,,
\eeq
where $\delta=\arctan(\epsilon/2\kappa)$.  The coupling constant 
becomes infinite at
\beq
  m  = \Lambda\exp\left[ -\frac1\kappa\left(\frac\pi2 + \delta\right)\right].
\eeq
We find that in the limit $N_c\to\infty$, the result from the RG
approach coincide with what is obtained from the Schwinger-Dyson
approach.  However, for finite $N_c$ the results of the two approaches
are different.  This is not unexpected, since the RG sums up a wider
class of diagrams than the gap equation.

For $g<g_*$, there are two zeros of the $\beta$-functions
\beq
  c_\pm = \frac1{N_c}\left(\pi\epsilon\mp \sqrt{\pi^2\epsilon^2-g^2N_c}\right).
\eeq
On the other hand, the scaling dimension of operator $\bar\psi\psi$ is
\beq
  \Delta_\pm[\bar\psi\psi] = 1+\epsilon - \frac{N_c}{2\pi} c_\pm
  = 1 + \frac\epsilon 2 \pm \sqrt{\pi^2\epsilon^2-g^2N_c}\,.
\eeq
We find that 
%at large $N_c$
\beq
  \Delta_+ + \Delta_- = 2+\epsilon = d\,.
\eeq
up to possible corrections of order $\epsilon^2$.
%Although there is possibly a $\epsilon^2$ correction in the right-hand
%side, we speculate that at large $N_c$ this result is exact in any
%dimensions $d$.  Indeed, when $\Delta_+=d/2$, the four-fermi coupling
%becomes exactly marginal, which should be expected if the $\beta$-function
%has a double zero.

\subsection{Summary of the relativistic example}

The lessons we learn from the relativistic examples are very similar
to the nonrelativistic example:
\begin{itemize}
\item $g<g_*$: two fixed points.  The dimensions of the operator 
$\bar\psi\psi$ at the two fixed points satisfy $\Delta_++\Delta_-=d$.
\item $g>g_*$: no fixed points, gap formation, BKT scaling.
\end{itemize}

\section{QCD at large $N_c$ and $N_f$.}
\label{sec:qcd}

We now turn our attention to the most interesting, and most difficult,
example; the chiral phase transition in QCD with large $N_c$ and
$N_f$.  Denote $x=N_f/N_c$.  We consider the Veneziano limit $N_c\to\infty$,
$N_f\to\infty$, $x$ fixed, and denote the (rescaled) 't Hooft coupling
as
\begin{equation}
  a_s = \frac{g^2N_c}{(4\pi)^2}\,.
\end{equation}
The beta function of QCD in this regime is
\begin{equation}
  \beta(a_s) = -\frac23 \left[ (11-2x)a_s^2 + (34-13x)a_s^3 + \cdots\right].
\end{equation}
For $x<11/2$ the theory is asymptotically free.  
If $x$ is slightly below $11/2$ the beta function has a 
nontrivial zero which is still in the perturbative regime:
\begin{equation}
 a_{s*} = \frac{2}{75}\left({11} - 2x\right).
\end{equation}
This is the Banks-Zaks (BZ) fixed point~\cite{Banks:1981nn}.  In the
IR, the theory is an interacting CFT.  This fixed point moves to
strong coupling as one makes $11/2-x\sim 1$.  For small $x$, $x\ll 1$,
we believe that the theory has chiral symmetry breaking and a
confinement scale.  It is natural to assume that there is a critical
$N_f/N_c$ ratio $x_{\rm crit}$ at which the chiral condensate goes to
zero.\footnote{One assumes that some UV scale, e.g., the scale
$\Lambda_{\rm QCD}$ defined from one-loop running, is fixed when $x$
is changed.}

If the picture emerging from the previous examples also holds for QCD,
then conformality is lost when the BZ fixed point annihilates with
another UV fixed point.  Therefore, we predict that when $x$ is
slightly larger than $x_{\rm crit}$, QCD has an UV fixed point, in
addition to the IR fixed point (and the free UV fixed point).

\begin{figure}[ht]
\includegraphics[width=0.6\textwidth]{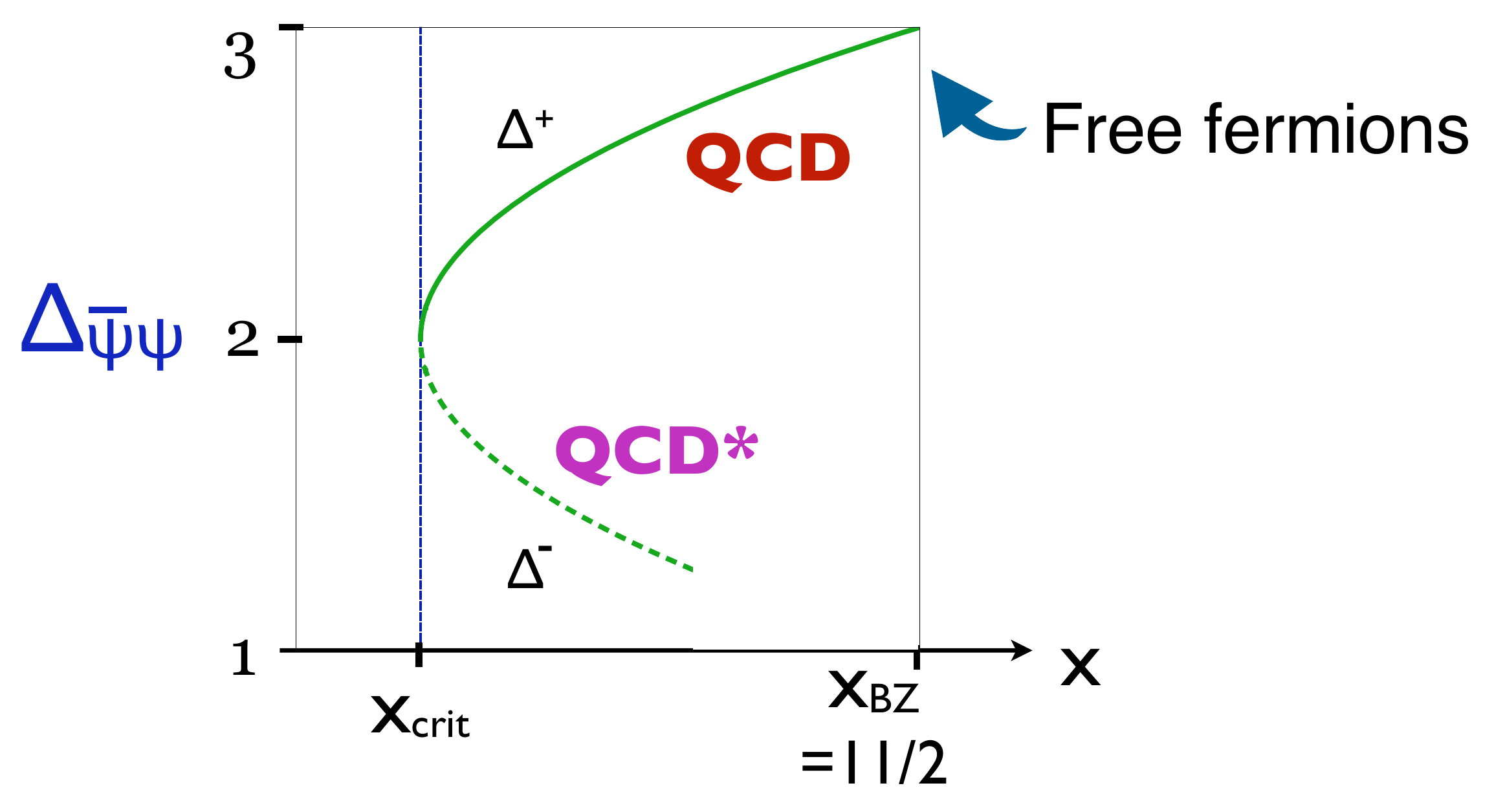}
\caption{A possible picture for the QCD chiral phase transition in $N_f/N_c$.
The lines denotes the dependence of the dimension of the chiral
condensate $\bar\psi\psi$ at the fixed point.  The solid line is the
IR fixed point, and the dashed line is the UV fixed point.}
\label{fig:qcd}
\end{figure}

The situation is illustrated in Fig.~\ref{fig:qcd}.  The UV fixed
point called QCD$^*$, is a different CFT compared to the usual IR
fixed point; for example, the dimension of the operator $\bar\psi\psi$
should be different between the two fixed points.

What is the nature of QCD$^*$?  It could be that the $\beta$-function
for the QCD gauge coupling simply has an unstable fixed point at
strong coupling.  However, this picture implicitly assumes that the
set of relevant operators in QCD$^*$ consist of just kinetic terms for
the gauge fields and fermions, as is the case at weak coupling.  At
strong coupling other operators could be relevant as well, and guided
by our defect QFT example of Sec.~\ref{sec:defect}, it is natural to
consider the possibility that a chirally symmetric four-fermion
operator is relevant in QCD$^*$
\begin{equation}\label{L-rough}
  {\cal L}_{\textrm{QCD}^*} = {\cal L}_{\textrm{QCD}}
    - c (\bar\psi \gamma^\mu t^a\psi)^2
\end{equation}
and that the unstable fixed point exists at some value $\{g_*, c_*\}$
in the two-dimensional space of couplings.  By analogy with
Sec.~\ref{sec:defect}, we then expect that the beta function for $c$
contains linear, quadratic and constant terms,
\begin{equation}
  \beta(c) = \gamma_1 c -  \gamma_2 N_c c^2 + \gamma_0 g^2,
\end{equation}
where the linear $\gamma_1 c$ term is due to the anomalous dimension
of the four-fermi operator, the quadratic $c^2$ term is due to the
one-loop graph involving two four-fermi vertices, and the constant
$g^2$ term is due to, e.g., one-gluon exchange graph.\footnote{The
constant term can be of a different power of $g$, but it does not
affect the argument.}  This is essentially
the picture advocated by Gies and 
Jaeckel~\cite{Gies:2005as}.\footnote{%
%A related work for finite temperature is \cite{Braun:2006jd}.  
Beta
functions containing three terms of the almost the same origin arise
in orbifolds of ${\cal N}=4$ super-Yang-Mills theory for the
coefficients of double-traced operators~\cite{Dymarsky:2005uh}.  The
constant term also appears in the running for the Landau liquid
parameters in the RG treatment of color
superconductivity~\cite{Son:1998uk}.} The constants $\gamma_0$,
$\gamma_1$, $\gamma_2$ depend on $x=N_f/N_c$; and $x_{\rm crit}$ is
where $\beta(c)$ has a double zero.

We do not know where in $x$ the fixed point QCD$^*$ exists.  A
particularly interesting possibility is that QCD$^*$ exists at weak
coupling, say near $x=11/2$.  As described subsequently, we find many
theories similar to QCD$^*$ in the perturbative regime, but none of
them possess the full chiral symmetry of QCD, and hence cannot be
QCD$^*$.  The possible ``phase diagram'' is illustrated in
Fig.~\ref{fig:qcd}, the line corresponding to QCD$^*$ does not
continue to the vicinity of $x=11/2$.

In the rest of this section, we shall be looking for perturbative UV
fixed points that flow to the BZ fixed point.  Models A and C below
have been considered in Ref.~\cite{Terao:2007jm}.

\subsection{Model A}

As the first step, we try to find a fixed point when \emph{one} operator,
$\bar\psi\psi$, has dimension different from its dimension at the BZ
fixed point.  From AdS/CFT experience, we expect that
operator dimensions at the two fixed points satisfy
\begin{equation}\label{sumDelta-modelA}
  \Delta_+ + \Delta_- = d = 4\,.
\end{equation}
At the BZ fixed point $\Delta_+\approx 3$, therefore at the QCD$^*$
fixed point, $\Delta_-\approx1$, which means that this operator is
almost a free scalar.  This suggests that we look for QCD$^*$ in the
following Lagrangian, which will be called model A,
\beq
    {\cal L}_{\textrm{model A}} = {\cal L}_{\rm QCD} + \frac12(\d_\mu\phi)^2
        - \frac y{\sqrt{2}}\bar\psi\psi\phi
        - \frac\lambda{24}\phi^4\,.
    \label{L-modelA}
\eeq
It is convenient to define the rescaled couplings,
\begin{equation}
  a_s = \frac{g^2N_c}{(4\pi)^2}\,,\quad
  a_y = \frac{y^2N_c N_f}{(4\pi)^2}\,,\quad
  \hat\lambda = \frac{\lambda N_c N_f}{(4\pi)^2}\,.
\end{equation}
These constants will be found to be $O(N_c^0)$ at the fixed point.  
In this regime, the
beta functions are
\begin{align}
  \beta_{a_s} &= - \frac23\bigl[(11-2x)a_s^2+(34-13x)a_s^3\bigr]\,,\\
  \beta_{a_y} & = - 6 a_s a_y + 2a_y^2,\\
%  \beta_{\hat\lambda} &= -\frac{a_y^2}2 + 2 a_y\hat\lambda\,,
  \beta_{\hat\lambda} &=  - 12 a_y^2 + 4 a_y \hat\lambda
\end{align}
The fixed point for the gauge coupling, $a_{s*}$, is the same as at
the BZ fixed point to leading order in $1/N_c$.  The fixed point for
the Yukawa and four-scalar couplings are
\begin{equation}
  a_{y*} = 3a_{s*}\,, \qquad \hat\lambda= 3a_{y*}= 9 a_{s*}\,.
\end{equation}
Thus, model A has a perturbative fixed point.

We can compute the dimension of the scalar operators at the fixed point,
\begin{subequations}
\begin{align}
  \Delta[\bar\psi\psi]_{\rm BZ} &= 3 - 3 a_{s*},\label{psipsimodelA} \\
  \Delta[\phi]_{\textrm{model A}} &= 1 + a_{y*}\,.
\end{align}
\end{subequations}
Notice that at in the model A fixed point, the operator $\bar\psi\psi$ is
replaced by the operator $\phi$.  We see here that
\beq
  \Delta[\bar\psi\psi]_{\rm BZ} + \Delta[\phi]_{\textrm{model A}} = 4\,,
\eeq
which coincides with our expectation~(\ref{sumDelta-modelA}). 
From the QM intuition, we may expect that when $\Delta_+=\Delta_-=2$,
the BZ fixed point and model A become identical.

According to our expectation, the new fixed point should be an UV
fixed point, and that there exist a deformation of this fixed point
that leads to the BZ fixed point.  The deformation is provided by the
scalar mass term $m^2\phi^2$.  This deformation is relevant if
$\Delta[\phi]$ is less than 2.  If such perturbation is present, the
scalar $\phi$ decouples below a certain energy scale, leaving the
theory to be in the BZ fixed point, as expected.

It might seem that the way model-A Lagrangian was introduced, with an
extra scalar field and Yukawa interaction, is very different from the
way done in Eq.~(\ref{L-rough}).  It seems that if we want to change
the dimension of $\bar\psi\psi$, then one should introduce a
four-fermi interaction into the QCD Lagrangian:
\begin{equation}\label{L-modelA2}
  {\cal L}_{\textrm{model A}} = {\cal L}_{\textrm{QCD}} + c(\bar\psi\psi)^2.
\end{equation}
We argue here, however, that the two forms of the Lagrangian are just
two different representations of the same fixed point; with
Eq.~(\ref{L-modelA}) being the weak-coupling representation near the
upper end of the conformal window ($x=11/2$), and
Eq.~(\ref{L-modelA2}) being the more useful representation near the
lower end ($x=x_{\rm crit}$).  Indeed, the propagator of $\phi$ at the
IR fixed point is $q^{2\Delta_\phi-4}$ where $\Delta_\phi$ is the
dimension of $\phi$; near the lower end of the conformal window
$\Delta=2$, therefore the scalar propagator is almost momentum
independent, and the scalar-mediated interaction between fermions 
becomes point-like.  This is similar to the equivalence between 
Nambu--Jona-Lasinio and Yukawa models in dimensions between 
2 and 4~\cite{Wilson:1972cf,ZinnJustin:1991yn}.
%(Here we are
%inspired by unitarity fermions, which can be represented as a fermions
%with four-point interaction or as a theory with fermions interacting with
%a dimer scalar field.)

\subsection{Model B}

Model B is similar to model A, except there are now two scalar fields,
\begin{equation}
        {\cal L}={\cal L}_{\rm QCD} 
        + \frac12(\d_\mu\phi_1)^2 + \frac12(\d_\mu\phi_2)^2
        - \frac y{\sqrt{2}}\bar\psi(\phi_1+i\gamma^5\phi_2)\psi  
        - \frac\lambda{24} (\phi_1^2+\phi_2^2)^2\,.
\end{equation}
The Lagrangian preserves vector SU($N_f$) and axial U(1)$_A$ (more
precisely, the nonanomalous discrete subgroup of it).  The behavior of
model B is exactly like in model A: there is a fixed point for $y$ and
$\lambda$; and the running of $g$ is not altered in large $N_c$, $N_f$
regime.

\subsection{Model C}

Both model A and B preserves only a small subset of the
SU($N_f$)$\times$SU($N_f$) chiral symmetry of QCD.  The simplest
Lagrangian which preserves chiral symmetry is
\beq
  {\cal L} = {\cal L}_{\rm QCD} - y ( \bar\psi t^A \psi \phi^A 
      + i \bar\psi t^A\gamma^5\psi \pi^A)
      + \Tr \d^\mu\Phi^\+ \d_\mu \Phi 
      - \lambda_1 (\Tr \Phi^\+\Phi)^2 - \lambda_2 \Tr (\Phi^+\Phi)^2\,,
\eeq
where $\Phi=(\phi^A+i\pi^A)t^A$, $A=0,\ldots N_f^2$ are flavor
Gell-Mann matrices, normalized so that
$\Tr(t^At^B)=\frac12\delta^{AB}$.  We need to find the fixed point of
this theory.  This fixed point should have only one IR unstable
direction corresponding to the mass term for $\Phi$.

The RG equations for $g$ and $y$ can be read out from 
Refs.~\cite{Machacek:1983tz,Machacek:1983fi,Machacek:1984zw}.
%\begin{align}
%  (4\pi)^2\frac{\d g}{\d s} &= \left(\frac{11}3 N_c - \frac23 N_f\right)g^3
%   + \left(\frac{34}3N_c^2-\frac{13}3 N_c N_f\right) \frac{g^5}{(4\pi)^2} 
%   + \frac1{(4\pi)^2}N_f^2 g^3y^2\\
%  (4\pi)^2\frac{\d y}{\d s} &= 3g^2 N_c y - (N_c+N_f) y^3
%\end{align}
%(we have assumed large $N_c$ and large $N_f$).  
We need the two-loop beta function for the gauge coupling (as the
one-loop contribution has a small coefficient near $x=11/2$), but for
the Yukawa and scalar couplings one-loop beta function suffices.
Moreover, the two-loop beta function for $g$ and the one-loop beta
function for $y$ does not contain scalar self-couplings, thus one can
first solve for the fixed point for $g$ and $y$, and then look for the
fixed points of scalar couplings.  In this model we define $a_y$ as
\beq
  a_y = \frac{y^2N_c}{(4\pi)^2}\,.
\eeq
The beta functions are
\begin{align}
  \beta_{a_s} &= -\frac23(11-2x)a_s^2
  - \frac23 (34-13x) a_s^3 - 2 x^2 a_s^2 a_y\,,\\
  \beta_{a_y} &= -6 a_s a_y + 2(1+x) a_y^2\,.
\end{align}
We work around $x=11/2$.  At fixed $a_s$, there is a zero of
$\beta_y$, but there is no fixed point of both beta functions.
Therefore model C does not have a perturbative fixed point.

\subsection{Model D}

The difference between model C and models A, B is that we introduce
$O(N^2)$ scalars into model C, while there are only $O(1)$ scalars in
models A, B.  As a result, the beta function for the gauge coupling
changes, and there is no longer a fixed point.

Our last model, model D, interpolates between models B and C.  We
introduce couplings to $2M^2$ scalars that preserve a
SU($M$)$\times$SU($M$)$\times$SU($k$) subgroup of the chiral symmetry
group, with $M=N_f/k$,
\begin{equation}
  {\cal L} = {\cal L}_{\rm QCD} -y \bar\psi^\alpha_i t^A_{\alpha\beta}
  (\phi^A + i\gamma^5\pi^A)\psi^\beta_i 
  + \textrm{scalar terms},
\end{equation}
where $\alpha$, $\beta$ runs $1\ldots M$, $i$ runs $1\ldots k$, 
$A$ runs $1\ldots M^2$.
Model B corresponds to $k= N_f$ and model $C$
to $k=1$.  We redefine $a_y$ to be
\begin{equation}
   a_y = \frac{y^2kN_c}{(4\pi)^2}\,.
\end{equation}
The beta functions are now
\begin{align}
  \beta_{a_s} &= -2a_s^2 \left[ \frac{11-2x}3 + \frac{34-13x}3 a_s 
     + \frac{x^2}{k^2} a_y\right].\\
  \beta_{a_y} &= 2a_y \left[ -3a_s + \left( 1+ 
    \frac x{k^2}\right)a_y\right].
\end{align}
For $x$ slightly below $11/2$, the model has a fixed point for 
any integer $k$ larger than 1,
\begin{equation}
  a_{s*}
    = \frac{2k^2+11}{25k^2-44} \cdot \frac{11-2x}{3}\,,\quad
  a_{y*}  = \frac{2k^2}{25k^2-44} (11-2x)\,,
\end{equation}
but there is no perturbative fixed point or $k=1$ (model C).  (The
fixed point values for the four-scalar couplings can also be found.)

Therefore, there exist theories that preserve part of the chiral
symmetry of QCD and flow to QCD by a relevant deformation, but we have
not succeeded in finding a theory that plays the role of QCD$^*$ in
the perturbative regime.  This does not mean QCD$^*$ does not exist;
%if it does, it has to do so 
in fact we will give arguments, largely based on holography, that it 
does exist in the nonperturbative region.

\subsubsection*{Operator dimensions in model D.}

The dimension of the scalar operators $\phi^A$, $\pi^A$ in model D is
\beq
  \Delta[\phi]|_{\textrm{model D}} = 1+ a_{y*}|_{\textrm{model D}}\,.
\eeq
On the other hand the dimension of $\bar\psi\psi$ at the BZ fixed
point is given in Eq.~(\ref{psipsimodelA}).  Taking the sum of the
dimensions, we find
\beq
  \Delta[\bar\psi\psi]|_{\textrm{BZ}} +\Delta[\phi]|_{\textrm{model D}}
  = 4 + \frac{88}{25(25k^2-44)}(11-2x)\,.
\eeq
So, the rule $\Delta_++\Delta_-=4$ is broken in model D when $k\sim
O(1)$.  When $k\gg1$, the equation  can be written in the suggestive form
\beq
  \Delta_+ + \Delta_- = 4 + \frac{88}{625}\frac{n_\phi}{N_f^2}(11-2x)\,,
\eeq
where $n_\phi=2M^2$ is the number of scalars.  We see that the
violation of the rule $\Delta_++\Delta_-=4$ occurs when the number of
scalars is of the same order as the number of color degrees of
freedom, $N_c^2$.

Recall that the rule $\Delta_++\Delta_-=4$ can be understood from
AdS/CFT correspondence: $\Delta$ is related to the mass square $m^2$
of the bulk scalar by the equations $\Delta(\Delta-d) = m^2R^2$.  How
do we understand the fact that this rule is violated when there are
$O(N^2)$ scalars?  In fact, it is easy to come up with a mechanism
leading to this effect within holography.  Recall the AdS radius $R$
is determined by the cosmological constant.  Changing the boundary
condition for the scalar field alters the vacuum energy (Casimir
energy) associated with that scalar field~\cite{Gubser:2002zh}.  The
change in the vacuum energy is small for one scalar, but becomes of
order one for $O(N^2)$ scalars.  Thus we have
\begin{equation}
  \Delta_\pm = \frac d2 \pm \sqrt{\frac{d^2}4 + m^2 R_\pm^2}
\end{equation}
and, if $R_+/R_-=1+O(n_\phi/N^2)$, then
$\Delta_++\Delta_-=4+O(n_\phi/N_c^2)$, which is exactly what is found.
We can construct an holographic model where the deviation of
$\Delta_++\Delta_-$ from 4 can be explicitly computed (see Appendix).
Interestingly, in this simple model $\Delta_++\Delta_->4$, as in model
D.

The holographic model also lends support to the hypothesis that model
C, with full chiral symmetry, exists in strong coupling.  Indeed, the
reason model C does not exist at weak coupling, from the holographic
point of view, is that flipping the boundary condition for $2N_f^2$
scalars from close to $z^3$ to close to $z^1$ is too much a disruption
for the AdS geometry (for example, in terms of the change of the
cosmological constant).  However, when both $\Delta_+$ and $\Delta_-$
are close to 2, the change of the vacuum energy is parametrically
small in $\Delta_+-\Delta_-$ (see Eq.~(\ref{eq:V-V})), and flipping
the boundary condition from $z^{\Delta_+}$ to $z^{\Delta_-}$ is no
longer a large disruption.  Hence, the theory where all fermion
bilinears have dimension $\Delta_-$, i.e., QCD$^*$, should exist near
the merger point.  However, arguments based on holographic models
can only be taken as suggestive at this moment.

\section{Conclusion}
\label{sec:concl}

There have recently been several lattice studies seeking to find the
boundaries of the conformal window in QCD
\cite{Appelquist:2007hu,Appelquist:2009ty,Deuzeman:2008da,Deuzeman:2009mh} 
and in other QCD-like theories
\cite{Sannino:2004qp,Dietrich:2006cm,Catterall:2007yx,Catterall:2008qk,Shamir:2008pb,Fodor:2008hm,DeGrand:2008kx,Hietanen:2008mr}. 
Interest in the
phase transition between conformal and nonconformal theories is
motivated in part by the invocation of approximate conformal symmetry
in numerous theories for physics beyond the Standard Model.

In this paper we have investigated the nature of such a phase transition, 
and we suggest that there
is a wide class of theories where it is due to the merger
and annihilation of fixed points. Several explicit examples 
of this phenomenon were given, 
and we speculate that
this mechanism is also responsible for the chiral phase transition in
QCD in the large $N_c$, large $N_f$ regime, at some critical value for
$N_f/N_c$.  We show that this mechanism leads to the BKT scaling
behavior of the chiral condensate at the phase transition, and also implies
the existence of the conformal theory QCD$^*$ which annihilates with 
QCD at the lower end of the conformal window. 
We tried, unsuccessfully, to construct QCD$^*$ in the perturbative regime,
and argued that it should exist in the nonperturbative regime.  It
would be interesting to search for evidence for QCD* on the lattice.

\begin{comment}
Beside the mechanism considered in this paper, conformality might also
be lost in other ways.  For example, consider the transition at the
upper end of the conformal window, $x=11/2$.  Above this value of $x$,
the trivial fixed point $g=0$ becomes stable, and the theory is in the
free electric Coulomb phase, where the coupling runs logarithmically
to zero in the IR, like in massless QED.  The same mechanism works in
supersymmetric QCD at $N_f/N_c=3$.

\begin{figure}[ht]
\includegraphics[width=0.4\textwidth]{seiberg}
\caption{Conformality may be lost if the fixed point runs off to infinity.}
\label{fig:seiberg}
\end{figure}

Anther possibility is that the fixed point runs off to infinity
(Fig.~\ref{fig:seiberg}).  This case may be the one realized in
supersymmetric QCD at $N_f/N_c=3/2$~\cite{Seiberg:1994pq}.  Here the
electric coupling goes to $\infty$, and the magnetic coupling goes to
$0$ logarithmically.  Of course, when the coupling is large, the beta
function is scheme dependent.  Nevertheless, we think this is a
qualitatively correct interpretation of the physics near this
transition.  For example, the slope of the beta function at its zero,
which is scheme-independent, approaches zero as $N_f/N_c$ approaches
$3/2$ from above~\cite{Anselmi:1996mq}.
\end{comment}

The models considered in the last section of our paper, in an attempt
to find the UV fixed point of QCD, may be of interest in their own
right.  For example, these models may be used to explicitly realize
the ``unhiggs''~\cite{Stancato:2008mp}, which behaves at high
energies as a field with noninteger scaling dimension.

The picture of the chiral phase transition realizes walking
technicolor when $N_f/N_c$ is only slightly below the phase
transition.  In the holographic interpretation, conformality is lost
when the mass squared of a bulk scalar drops below the BF bound.  A
naive application of AdS/CFT rules implies that the dimension of the
operator $\bar\psi\psi$ is equal to 2 at the phase transition.  This 
feature is explicit in the holographic model considered in the Appendix.
This conclusion is,
interestingly, in agreement with result from the Schwinger-Dyson
approach, and also with Ref.~\cite{Cohen:1988sq}.  The result
illustrates that the dimension of the fermion bilinear on the IR
stable branch cannot approach the unitarity bound used in
Ref.~\cite{Ryttov:2007cx} for estimating the maximal extension of the
conformal window.

\acknowledgments

The authors thank Ken Intriligator, Andreas Karch, and Igor Klebanov
for discussions, and Brian Mattern for comments on the manuscript.  
This work is supported, in part, by U.S.\ DOE grants
No.\ DE-FG02-00ER41132 (DBK, JWL, DTS) and DE-FG02-01ER41195 (MAS).

\appendix
\section{Casimir effect in AdS$_5$}

We consider a holographic model for a UV-IR pair of conformal
theories.  Each theory has a set of $n_\phi$ scalar operators that
all have the scaling dimension
$1<\Delta_-<2$ in one theory and $2<\Delta_+<3$ in the
other. 
Both theories are described by the same holographic dual
  \begin{equation}\label{eq:S5}
\begin{split}
S_5 = \frac{1}{2 \kappa^2}&\int_M \!\! d^5x \; \sqrt{-g}\,
\biggl( {\cal R} - V_0 
- \frac{1}{2}\sum_{i=1}^{n_\phi}\left( 
(\partial \phi_i)^2  + m^2 \phi_i^2\right)
\biggr)\,.\\
%& - 2 \int_{\partial M} \!\!\!\! d^4x \; \sqrt{-\gamma}\, K \biggr),
\end{split}
\end{equation}
The two different theories correspond to the two choices of the
boundary conditions on the scalar fields $\phi_i$. The solution of the
classical equations of motion is given by all $\phi_i=0$ and AdS$_5$ metric:
\begin{equation}
  \label{eq:ds2}
  ds^2= R_0^2\, z^{-2}\, (dz^2+d\bm{x}^2-dt^2)\,,
\end{equation}
with
\begin{equation}
  \label{eq:L0}
  R_0^{2} = -12/V_0\,.
\end{equation}
The loop expansion is controlled by dimensionless parameter 
$\kappa^2 R_0^{-3}\sim N_c^2$,
where $N_c$ is the number of colors in the dual theory. One-loop
contribution is not negligible in the $N_c\to\infty$ limit if $n_\phi$
is also large, i.e., $n_\phi\sim N_c^2$.  For simplicity we assume 
$n_\phi/N_c^2\ll1$ and compute $\Delta_++\Delta_--4$ to leading order in 
$n_\phi/N_c^2$.  We use the technique of 
Ref.~\cite{Gubser:2002zh}.\footnote{For earlier calculations of the Casimir
energy in AdS space see Refs.~\cite{Garriga:2000jb,Goldberger:2000dv}.}
The one-loop contribution of
the scalar fields depends on the boundary condition. This contribution
shifts the vacuum energy $V_0$ to $V_\pm$. Consequently, the
AdS$_5$ curvature radius $\Ro$ becomes at one loop
\begin{equation}
  \label{eq:Lpm}
  R_\pm^2=-12/V_\pm \,.
\end{equation}
It is convenient to measure lengths in units of $\Ro$, i.e., $\Ro=1$, 
$V_0=-12$, etc.

The calculation of the vacuum energy is easier to perform after the
Wick rotation $t\to -ix_4$. The correction to the vacuum energy is equal to
\begin{equation}
  \label{eq:Vpm-V0}
  \int\!d^5x_E\sqrt{g_E}(V_\pm - V_0) = 2\kappa^2\,
\frac{n_\phi}2\,\log{\det}_\pm\left[(-\nabla_E^2+m^2)\sqrt{g_E}\right],
\end{equation}
where index $E$ denotes objects defined using the Wick rotated metric
$ds_E^2=(dz^2+d\bm{x}^2+dx_4^2)$. The expression in the r.h.s. of 
Eq.~(\ref{eq:Vpm})
 is formal,
since it is UV divergent. The derivative with respect to $m^2$ eliminates some
but not all divergences:
\begin{equation}
  \label{eq:Vpm}
  \int\!d^5x_E\sqrt{g_E}\,\frac d{dm^2}V_\pm= \kappa^2 \, n_\phi \,
{\rm tr_\pm}\,\left\{\left[(-\nabla_E^2+m^2)\sqrt{g_E}\right]^{-1}
\sqrt{g_E}\right\}.
\end{equation}
The kernel of the operator $\left[(-\nabla_E^2+m^2)\sqrt{g_E}\right]^{-1}$ 
is the
Green's function defined by the following equation
\begin{equation}
  \label{eq:Geq}
  \left[-\partial_z z^{-3} \partial_z + Q^2z^{-3} + m^2 z^{-5}\right]
G_{\pm\nu}(z,z';Q)=\delta(z-z')\,.
\end{equation}
where $\pm\nu$ refer to two different boundary conditions at $z=0$:
$G_{\pm\nu}\sim z^{2\pm\nu}$, where $\nu=\sqrt{4+m^2}$, and $Q^2=q_E^2$. The
solution to Eq.~(\ref{eq:Geq}) is 
%easy to find (using the Wronskian of Bessel functions):
\begin{equation}
  \label{eq:G-JK}
  G_{\pm\nu}(z,z';Q) = z^2\, z'^2\, 
  I_{\pm\nu}(Qz)\,K_{\pm\nu}(Qz')\, \theta(z'-z)+(z\leftrightarrow z')\,.
\end{equation}
Collecting results so far, we find
\begin{equation}
  \label{eq:Vpm-G}
  \frac d{dm^2}V_\pm= \kappa^2\,n_\phi \,\int\! 
  \frac{d^4q_E^{\phantom{1}}}{(2\pi)^4}\,
G_{\pm\nu}(z,z;Q)\,.
\end{equation}
The right hand side of Eq.~(\ref{eq:Vpm-G}) is still UV divergent. However, the
difference $V_+-V_-$ is finite:
\begin{equation}
  \begin{split}
    \label{eq:dV-dV)}
    \frac1{\kappa^2 n_\phi}\,\frac d{dm^2}(V_+-V_-)&=
\frac1{8\pi^2} \int_0^\infty\! dQ\, Q^3\,
    z^4 \left(
I_{\nu}(Qz)\,K_{\nu}(Qz)-(\nu\leftrightarrow-\nu)
\right) \\
    &= -\frac1{8\pi^2} \frac{2\sin\nu\pi}{\pi}
\int_0^\infty\!dx\, x^3\, K_\nu^2(x) = -\frac1{12\pi^2}\,\nu(1-\nu^2)\,.
  \end{split}
\end{equation}
Using the fact that both boundary conditions are the same at
$\nu=0$, and thus $V_+=V_-$ at $\nu=0$, as well as $d\nu^2=dm^2$, we
can write
\begin{equation}
  \label{eq:V-V}
 \frac1{\kappa^2 n_\phi}\, (V_+-V_-) = -\frac1{12\pi^2}\int_0^{\nu^2}\!
 d\tilde \nu^2 \,\tilde\nu(1-\tilde\nu^2)
= -\frac1{6\pi^2}\left(\frac{\nu^3}3-\frac{\nu^5}5\right)<0\,.
\end{equation}
(Note that $0<\nu<1$.)

Using Eqs.~(\ref{eq:L0}), (\ref{eq:Lpm}) and in the regime
$|\Rpm-\Ro|\ll \Ro=1$, we can write
\begin{equation}
  \label{eq:L-L}
  \Rp-\Rm=\frac16\,({V_+-V_-})\,.
\end{equation}
The one-loop corrected scaling dimensions of the scalar operators
become
\begin{equation}
  \label{eq:Dpm}
  \Delta_\pm=2\pm\sqrt{4+m^2 R_\pm^2}\,.
\end{equation}
and thus
\begin{equation}
  \label{eq:D+D}
  \Delta_++\Delta_--4 = \frac{m^2}{\nu}\,(\Rp-\Rm)
=
\frac{\kappa^2 n_\phi}{36\pi^2}\,\nu^2\,(4-\nu^2)\,
\left(\frac13-\frac{\nu^2}5\right) > 0\,.
\end{equation}

Since $\kappa^2\sim N_c^2$ the deviation of $\Delta_++\Delta_-$ from 4
is $O(n_\phi/N_c^2)$.  At the point of merger $\nu\to0$,
$\Delta_++\Delta_-=4$, hence $\Delta_+=\Delta_-=2$, with no correction
of order $O(n_\phi/N_c^2)$.

\bibliography{BKT}
\bibliographystyle{apsrev}

\end{document}